\documentclass[aps,pre,10pt,twocolumn,superscriptaddress,nolongbibliography]{revtex4-1}
\usepackage{amsmath}
\usepackage{amsfonts}
\usepackage{mathrsfs}
\usepackage{graphicx,color}
\usepackage{bm,bbm}
\usepackage{multirow}
\usepackage{hyperref}
\usepackage{physics}
\usepackage{enumitem}
\usepackage{cases}
\usepackage[caption=false]{subfig}
\usepackage{algorithm,algpseudocode}

\makeatletter
\newcommand\footnoteref[1]{\protected@xdef\@thefnmark{\ref{#1}}\@footnotemark}
\makeatother

\DeclareMathAlphabet{\mathdj}{U}{msb}{m}{n}

\newcommand{\OO}{\mathcal{O}}
\newcommand{\ee}{\mathrm{e}}

\begin{document}

\title{High-dimensional percolation criticality and hints of mean-field-like caging of the random Lorentz gas}
\author{Benoit Charbonneau}
\affiliation{Department of Pure Mathematics, University of Waterloo, Waterloo, Ontario N2L 3G3, Canada}
\affiliation{Department of Physics and Astronomy, University of Waterloo, Waterloo, Ontario N2L 3G3, Canada}

\author{Patrick Charbonneau}
\affiliation{Department of Chemistry, Duke University, Durham, North Carolina 27708, USA}
\affiliation{Department of Physics, Duke University, Durham, North Carolina 27708, USA}

\author{Yi Hu}
\email{yi.hu@duke.edu}
\affiliation{Department of Chemistry, Duke University, Durham, North Carolina 27708, USA}

\author{Zhen Yang}
\affiliation{Department of Physics, Duke University, Durham, North Carolina 27708, USA}
\affiliation{Kuang Yaming Honors School, Nanjing University, Nanjing 210023, China}

\date{\today}

\begin{abstract}
The random Lorentz gas (RLG) is a minimal model for transport in disordered media. 
Despite the broad relevance of the model, theoretical grasp over its properties remains weak. For instance, the scaling with dimension $d$ of its localization transition at the void percolation threshold is not well controlled analytically nor computationally. A recent study [Biroli \emph{et al}. \emph{Phys. Rev. E} L030104 (2021)] of the caging behavior of the RLG motivated by the mean-field theory of glasses has uncovered physical inconsistencies in that scaling that heighten the need for guidance. Here, we first extend analytical expectations for asymptotic high-$d$ bounds on the void percolation threshold, and then computationally evaluate  both the threshold and its criticality in various $d$. In high-$d$ systems, we observe that the standard percolation physics is complemented by a dynamical slowdown of the tracer dynamics reminiscent of mean-field caging. A simple modification of the RLG is found to bring the interplay between percolation and mean-field-like caging down to $d=3$. 
\end{abstract}
\maketitle

\section{Introduction} 

The broad reach of the simple percolation universality class~\cite{stauffer1994percolation,ben2000diffusion} enables the effective use of simple physical models to capture  features of systems as varied and complex as fractured geological formations~\cite{berkowitz2006modeling}, molecular transport in cells~\cite{hofling2013anomalous}, and epidemic spreading~\cite{dorogovtsev2008critical}. 
Percolation physics indeed unifies key aspects of critical clustering, conductance, and anomalous transport in heterogeneous media.
Yet not all critical aspects of percolation are universal. Exponents associated with conductance and transport, for instance, are not.  The value of the critical exponent for conductance and diffusion, $\mu$, is affected by the distribution of bond strengths in a random resistor network~\cite{straley1982non}, or equivalently, by the distribution channel widths in continuous space models~\cite{halperin1985differences}. Critical thresholds also strongly depend on microscopic details. As a result, certain key features of even simple percolation models remain poorly characterized.

In general, continuous-space percolation is less well understood than its lattice counterpart. While ever more accurate critical exponents and thresholds keep being reported for lattice percolation~\cite{xu2014simultaneous,mertens2018percolation,zhang2020critical}, inconsistent theoretical predictions about diffusion and subdiffusion critical exponents for dynamical processes in continuous space persisted for decades~\cite{machta1985diffusion,havlin1987diffusion}, before H{\"o}fling~\emph{et al.}~\cite{hofling2006localization,hofling2008critical} could validate the proposal of Machta~\emph{et al.}~in $d=3$~\cite{machta1985diffusion}. Similar inconsistencies for $d>3$ remain unexamined, and still sharper incongruities between descriptions of percolation in $d\rightarrow\infty$ continuous space have recently emerged~\cite{biroli2020unifying}. 

Before going any further, let us describe the specific continuous-space model of interest: $N$ spherical obstacles of radius $\sigma_\mathrm{obs}$ placed uniformly at random within a box of volume of $V$. A unitless density can then be defined,
\begin{equation} \label{eq:densitydef}
\Phi = \rho V_d \sigma_\mathrm{obs}^d,
\end{equation}
where  $\rho=N/V$ is the number density of spheres, and $V_d = \pi^{d/2}/\Gamma(1+d/2)$ is the $d$-dimensional volume of a ball of unit radius. The thermodynamic limit for this model consists of having $N$ and $V$ both diverge, while keeping $\rho$ (and thus $\Phi$) constant. By contrast to lattice models, which offer a simple duality between occupied and unoccupied sites, two nonequivalent types of percolation can here be identified: (i) the direct percolation of the overlapping sphere network; (ii) the percolation of the interstitial void (or vacant~\cite{Meester1996}) space. Despite the superficial similarities between the two phenomena, our understanding of them differs markedly. The asymptotic, high-dimensional scaling of the direct percolation threshold has been physically argued~\cite{torquato2012effect1}, rigorously proven~\cite{anantharam2016boolean}, and numerically assessed up to $d=11$~\cite{torquato2012effect2}. By contrast, reports of percolation thresholds for $d>3$ are limited~\cite{jin2015dimensional}, and the asymptotic high-dimensional scaling of that threshold remains an open question.

Another way of characterizing void percolation is to consider a point-like tracer, with radius $\sigma_\mathrm{tracer}=0$, traveling within that void space. (For a reason that will become obvious below, we set $\sigma=\sigma_\mathrm{obs}+\sigma_\mathrm{tracer}$ as the unit of length. In the rest of this paper, $r$ denotes unitless lengths, i.e., scaled by $\sigma$.) This transport process defines the random Lorentz gas (RLG), which can be construed as a minimal model of structural glasses.
As a result the RLG and its variants have been extensively studied by the mode-coupling theory of glasses~\cite{gotze1981dynamical,leutheusser1984dynamical,szamel2004gaussian,krakoviack2007mode,kim2009slow,kurzidim2009single,szamel2013glassy,jin2015dimensional} and, more recently, by the mean-field theory of glasses~\cite{biroli2020unifying,biroli2020mean}.
Although these descriptions are generally consistent with numerics away from the percolation threshold---both in the diffusive~\cite{kim2009slow,kurzidim2009single} and the localized~\cite{biroli2020unifying} regimes---percolation criticality is not well captured by either~\cite{jin2015dimensional,biroli2020unifying}.

A recent study suggests that in finite $d$, percolation criticality controls the long-time dynamics, whereas glassy dynamics intervenes at intermediate time scales~\cite{biroli2020unifying}. 
This scenario motivates a more detailed characterization of the intermediate-time dynamical slowdown, which although a mere pre-asymptotic feature of percolation theory, is fundamental to understanding finite-dimensional glass physics. As further motivation for considering this regime, we note that active colloids evolving within soft obstacles slow down at intermediate times but without affecting the long-time localization transition~\cite{morin2017diffusion}, and that replacing the hardcore by a Lennard-Jones-like repulsion retains the subdiffusion criticality~\cite{petersen2019anomalous}. In addition, although the glassy behavior of the RLG is only transient in finite $d$, it is conceivable that small modifications to the model could enhance its glass-like features, and thus strengthen the physical analogy with glasses. 

In this paper, we detail and use recent numerical advances to study the caging regime of the RLG and to evaluate its percolation threshold and criticality in high dimension, paying particular attention to the intermediate-time dynamical regime. The plan for the rest of this article is as follows. Section~\ref{sec:scaling} briefly reviews critical scaling relations for the RLG around the void percolation threshold, then
Sec.~\ref{sec:bounds} describes our conjecture for (loose) $d\rightarrow\infty$ asymptotic bounds for that threshold.
Section~\ref{sec:methods} details the computational schemes used to evaluate both the percolation threshold and some of its critical properties, and Sec.~\ref{sec:results} presents and discusses the associated numerical results.
In light of these findings, Sec.~\ref{sec:inhomo} proposes and studies a modified RLG that enhances the intermediate-time glass-like dynamical caging regime. We briefly conclude in Sec. ~\ref{sec:conclusion}.

\section{Percolation criticality and scaling analysis} \label{sec:scaling}

In this section we briefly review the main critical exponents that describe void percolation as well as scaling relations between them. Beforehand, note that although the volume fraction of void space, $\eta = e^{-\Phi}$, is formally equivalent to the covering fraction in lattice percolation and thus may seem to be a natural choice to describe the critical around the percolation threshold, $\eta_\mathrm{p} = e^{-\Phi_\mathrm{p}}$, numerical studies use $\Phi$, which is linearly related to $\eta$ to the lowest order,
\begin{equation}
\eta - \eta_\mathrm{p} = -\eta_\mathrm{p}\{\Phi - \Phi_\mathrm{p} + \OO[(\Phi - \Phi_\mathrm{p})^2] \},
\end{equation}
because it leads to less pronounced pre-asymptotic scaling corrections~\cite{hofling2008critical,bauer2010localization}. 
We thus here define the distance to the percolation threshold as $\epsilon=(\Phi - \Phi_\mathrm{p})/\Phi_\mathrm{p}$.

Upon approaching $\Phi_\mathrm{p}$, the long-time behavior of the mean squared displacement (MSD) of the tracer, $\Delta(t, \epsilon) = \langle r^2(t) \rangle$ is expected to scale as~\cite{ben2000diffusion}
\begin{subnumcases}{\Delta(t \rightarrow \infty, \epsilon) \sim}
|\epsilon|^{\mu_-}, & $\epsilon > 0$, \label{eq:localscaling}\\
t |\epsilon|^\mu, & $\epsilon < 0$, \label{eq:diffusionscaling}\\
t^{2/d'_\mathrm{w}}, & $\epsilon = 0$,\label{eq:subdiffusionscaling}
\end{subnumcases}
where $\mu_-$ is the localization exponent for the cage size ($\Delta(t \rightarrow \infty$)), $\mu$ is the diffusion exponent, and $d'_\mathrm{w} > 2$ is the subdiffusion exponent. Combining the critical scaling forms of the localization and of the diffusion regimes then leads to a scaling collapse of the MSD~\cite{ben2000diffusion},
\begin{equation} \label{eq:uniscale}
\Delta(t, \epsilon) = |\epsilon|^{- \mu_-} f[\mathrm{sgn}(\epsilon) |\epsilon|^{(\mu_- + \mu)} t ].
\end{equation}

For the localization and subdiffusion exponents, the critical scaling analysis further predicts that~\cite{stauffer1994percolation} 
\begin{equation} \label{eq:subdiffscale}
d'_\mathrm{w} = 2\left( \frac{\mu}{\mu_-} + 1\right).
\end{equation}
At the percolation threshold, the cluster size distribution, here given by the distribution of cavity volumes, $V_\mathrm{cavity}$, is thus expected to scale as
\begin{equation} \label{eq:fisherscaling}
P(V_\mathrm{cavity}; \epsilon=0) \sim V_\mathrm{cavity}^{-\tau}
\end{equation}
where $\tau$ is the Fisher exponent~\cite{stauffer1994percolation}.

Although geometric critical exponents, including $\tau$, $\mu_-$ and the correlation length exponent $\nu$ are universal within the simple percolation universality class, the conduction and transport exponent $\mu$ may differ from one model to another and may even depend on the specifics of the dynamics~\cite{halperin1985differences,spanner2016splitting}. The general form is~\cite{lubensky1986varepsilon,stenull2001conductivity}
\begin{equation} \label{eq:diffusionscalegeneral}
\mu =  \begin{cases}
\max[\mu_\mathrm{lattice}, \nu(d-2) + 1/(1-\alpha)], & d < d_\mathrm{u} \\
\max[3, 2 + 1/(1-\alpha)], & d \ge d_\mathrm{u} \\
\end{cases}
\end{equation}
where $d_\mathrm{u} = 6$ is the upper critical dimension for percolation, and
$0 < \alpha < 1$ is a model-dependent factor originating from the continuous distribution of bond strengths in the percolated cluster~\cite{straley1982non,stenull2001conductivity}.
For the RLG with ballistic dynamics, in particular, Ref.~\onlinecite{machta1985diffusion} predicts that $\alpha = (d-2)/(d-1)$, and thus
\begin{equation} \label{eq:diffusionscale}
\mu =  \begin{cases}
\mu_\mathrm{lattice}, & d = 2, \\
\nu(d-2) + d - 1, & 2 < d < 6, \\
d + 1, & d \ge 6. \\
\end{cases}
\end{equation}
This prediction is supported by numerical studies in $d=2$~\cite{bauer2010localization} and $d=3$~\cite{hofling2006localization,hofling2008critical}, but no numerical assessment exists for $d \ge 4$.
Interestingly, we also have that for $d \ge d_\mathrm{u}$, $\mu_-$ vanishes and $d'_\mathrm{w}$ diverges. The divergence of the cage size upon approaching $\Phi \rightarrow \Phi_\mathrm{p}^+$ and the long time limit of the subdiffusion  at $\Phi_\mathrm{p}$ then both scale logarithmically~\cite{biroli2019dynamics}.

\section{Percolation Threshold bound conjectures}
\label{sec:bounds}

In this section we first review known and then derive tighter asymptotic $d\rightarrow\infty$ scalings of both lower and upper bounds to the void (and thus RLG) percolation threshold. Note that for this system the probability that no obstacle center lies within a volume $\tilde{V}$ (of arbitrary shape) is given by $\ee^{-\rho \tilde{V}}$, because the randomly distributed obstacles form a Poisson point process. It immediately follows that: (i) the volume fraction of vacant space of obstacles is $\ee^{-\rho V_d \sigma_\mathrm{obs}^d} = e^{-\Phi r_\mathrm{obs}^d}$ and (ii) the volume fraction available to the tracer center is $\eta = \ee^{-\Phi (r_\mathrm{obs} + r_\mathrm{tracer})^d}$. 
Note also that in the original RLG setting, $r_\mathrm{obs}=1, r_\mathrm{tracer}=0$, which properly recovers $\eta = \ee^{-\Phi}$ from Sec.~\ref{sec:scaling}, but other equivalent choices are possible. 

\subsection{Prior work}

While the scaling of the direct (occupied space) percolation threshold is under analytical control, $\lim_{d\rightarrow\infty}\Phi_\mathrm{c} \sim 2^{-d}$~\cite{penrose1996continuum,anantharam2016boolean} (see also Refs.~\onlinecite{torquato2012effect1,torquato2012effect2}), little is formally known about the void (vacant space) percolation threshold, $\Phi_\mathrm{p}$, other than that $\Phi_\mathrm{c}$ provides a lower bound for it,
\begin{equation}
\Phi_\mathrm{p} = \Omega(\Phi_\mathrm{c}) = \Omega(2^{-d}).
\end{equation}
Numerical results for $d=2$ to 9~\cite{jin2015dimensional,biroli2020unifying}, however, suggest that this bound is very loose, with $\Phi_\mathrm{p}$ growing increasingly distant from $\Phi_\mathrm{c}$ as $d$ increases. 

No asymptotic upper bound is known. Asymptotic results for the volume fraction threshold for a Poisson point process, such as $\lim_{d\rightarrow\infty}\Phi_\mathrm{v}(d) = \OO(1)$~\cite{anantharam2016boolean}, do not help. Even when all points of space are covered by an obstacle with probability one, a tenuous pathway of voids can still percolate. The percolation universality class (in physics) indeed suggests that at $\Phi_\mathrm{p}$ the percolating cluster is a giant component with fractal dimension $d_\mathrm{f} = 4$ for $d \ge d_\mathrm{u} = 6$, and hence the volume fraction of vacant space at that threshold is expected to vanish in the limit of $d \rightarrow \infty$. Unsurprisingly, numerical results strongly suggest that $\Phi_\mathrm{p}(d)\gtrsim\OO(d)\gg\Phi_\mathrm{v}(d)$. 

\subsection{Upper bound} \label{sec:phipup}

We here propose a physically motivated upper bound for $\Phi_\mathrm{p}$. First, we note that the obstacle and tracer radii can be changed---as long as their sum remains $\sigma$---without changing the volume fraction of space available to the tracer center, $\ee^{-\Phi}$. Consider now a tracer whose center lies in this available space and, for convenience, defines the origin. In other words, within a ball of radius $r_\mathrm{tracer} + r_\mathrm{obs} \equiv 1$ around the origin, no obstacle center can be found. 
The volume fraction of space vacant of obstacles, however, does then vary as $\ee^{-\Phi r_\mathrm{obs}^d}$. 
We wish to find an asymptotic bound, $\Phi_\mathrm{u}(d)$, such that a potential pathway from the origin to a far away point, i.e., a percolating path, is infinitely suppressed in the limit $d\rightarrow\infty$. Because the contrapositive statement, \emph{the probability that such a pathway exists vanishes} is a sufficient condition for void space then not to percolate, we have that $\Phi_\mathrm{u}(d)$ must be an upper bound for the percolation threshold. 	

From percolation theory, we know that for $\Phi \rightarrow \Phi_\mathrm{p}^{-}$  the fraction of vacant space belonging to the (unique) percolated cluster scales as $(1 - \Phi/\Phi_\mathrm{p})^\beta$ with $\beta=1$ for $d \ge d_\mathrm{u}=6$~\cite{stauffer1994percolation}. Hence, if a percolating cluster exists, i.e., for $\Phi < \Phi_\mathrm{p}$, then the probability that a random uncovered point belongs to the infinite percolated cluster of void space is \emph{finite}. Because this critical scaling is physically motivated rather than a mathematical theorem, however, the following demonstration remains but a conjecture.

\begin{figure*}
\includegraphics[width=0.9\textwidth]{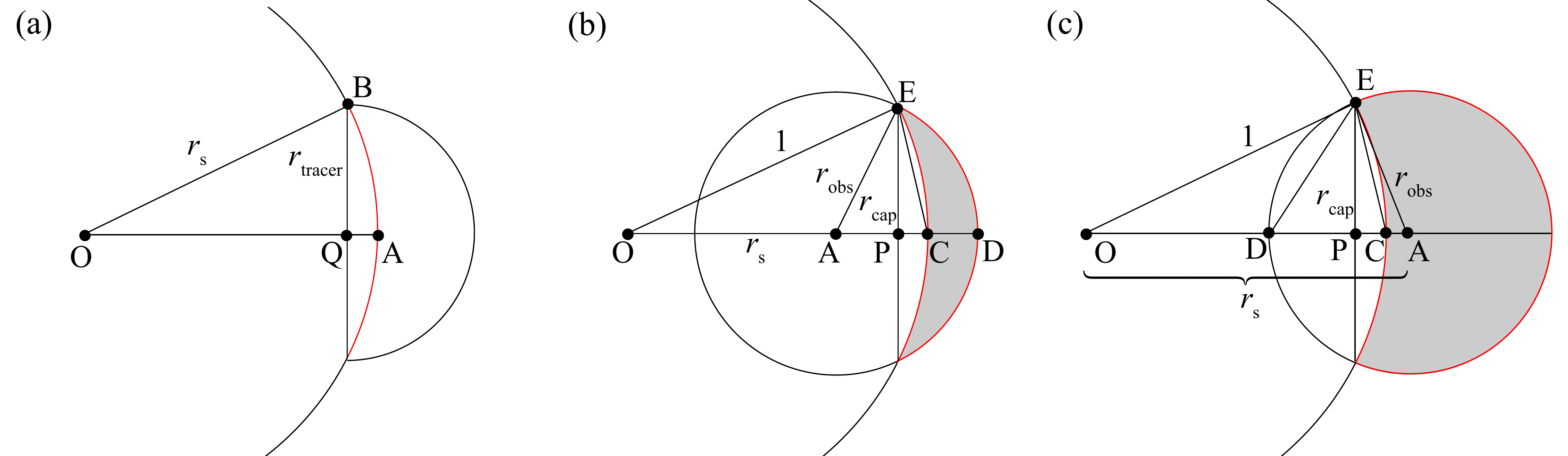}
\caption{Spherical cap geometry. (a) A tracer initially situated at the origin O passes through a shell of radius $r_\mathrm{s}$ = OB as its center moves to point Q with $r_\mathrm{tracer}$ = QB. Because the tracer center fits at Q, a spherical cap of sufficient area (red) must be devoid of obstacles.
(b, c) Point A on the spherical shell of radius $r_\mathrm{s}$=OA lies within the void space if no obstacle center appears in the shaded area (Because the origin is available for the tracer center, no obstacle center can lie within a radius $r$=OE=1 from the origin.) which is the set minus of a sphere centered on A of radius AE = $r_\mathrm{obs}$ to the unit sphere centered at origin of radius OE = 1 (E denotes a point on the intersection of these two spheres). In (b) the spherical caps on sphere OE and AE are on the same side of their common base, and in (c) they are on different sides. In both cases, these spherical caps have base radius $r_\mathrm{cap}=\mathrm{PE}$.
}
\label{fig:geometry}
\end{figure*}

To enable the tracer to leave a spherical shell situated a distance $r_\mathrm{s}$ from the origin, there must exist at least one sufficiently large hole within that shell. Denoting $A_\mathrm{s}$ the expected area on this shell that lies within the vacant space, and $A_\mathrm{h}$ the area of the intersection between the shell surface and the tracer, the probability that a shell has such a hole is then bounded by 
\begin{equation} \label{eq:Pholeexist}
P_\mathrm{h} \le \min(1, A_\mathrm{s}/A_\mathrm{h}).
\end{equation}
As can be seen in Fig.~\ref{fig:geometry}(a), $A_\mathrm{h}$ depends on both $r_\mathrm{s}$ and $r_\mathrm{tracer}$ as long as $r_\mathrm{s} > r_\mathrm{tracer}$.
From Fig.~\ref{fig:geometry}(b,c), we see that $A_\mathrm{s}$ also depends on $r_\mathrm{s}$ and $r_\mathrm{obs}$, as well as on $\Phi$. 
For $A_\mathrm{s} > A_\mathrm{h}$, Eq.~\eqref{eq:Pholeexist} reduces to $P_\mathrm{h} \le 1$ and thus contains no useful information about the tracer escaping the shell. For example, when $r_\mathrm{tracer} = 0$ we have $A_\mathrm{h} = 0$. Hence this construction fails under the standard formulation of the RLG with an infinitely small tracer.
Adjusting $r_\mathrm{obs} \in (0, 1)$ keep the model equivalent to the RLG, but $A_\mathrm{h}$  then takes a nonzero finite value as long as $r_\mathrm{s} > r_\mathrm{tracer} = 1 - r_\mathrm{obs}$. Because $A_\mathrm{s}$ vanishes as $\Phi \rightarrow \infty$, there exists a dimensional scaling of $\Phi \sim f_\mathrm{u}(d)$ for $A_\mathrm{s}/A_\mathrm{h}$ that vanishes in the limit $d \rightarrow \infty$. Here, we specifically seek the largest lower bound on $f_\mathrm{u}(d)$ among all valid choices of $(r_\mathrm{s}$, $r_\mathrm{obs})$, namely,
\begin{equation} \label{eq:infbound}
\Phi_\mathrm{u}(d) = \inf_{(r_\mathrm{s}, r_\mathrm{obs})} f_\mathrm{u}(d), \quad r_\mathrm{s} > 1 - r_\mathrm{obs} \text{ and } 0 < r_\mathrm{obs} < 1.
\end{equation}
The $\Phi_\mathrm{u}(d)$ that gives the smallest upper bound on $\Phi_\mathrm{p}$ can then be obtained by an asymptotic calculation, as detailed in Appendix~\ref{appd:phipup},
\begin{equation} \label{eq:phipuptheory}
\Phi_\mathrm{p}(d) \le \Phi_\mathrm{u}(d) = \OO(d \log d).
\end{equation} 
Note, however, that this upper bound may not be tight because the inequality in Eq.~\eqref{eq:Pholeexist} may itself not be tight.

\subsection{Lower bound} \label{sec:phiplw}

Physically-motivated improvements can also be made to the lower bound by noting that a simple sufficient condition for a $d$-dimensional system to percolate is that an $n$-dimensional subsystem, with $n < d$, should also percolate.  Obstacles in this low-dimensional subsystem are cross-sections of $d$-dimensional balls of unit radius with an $n$-dimensional hyperplane and are thus randomly distributed obstacles with unitless radii $r_\mathrm{obs} \in (0, 1]$. (Note that this definition differs from that in Sec.~\ref{sec:phipup}.) 
Once the number density and radius distribution of these obstacles are known, then we can examine if the vacant space in the subsystem percolates.

For simplicity we consider $n \sim \OO(1)$. The number of obstacle centers on this hyperplane is then the expected number of obstacles in the $d$-dimensional system intersecting this plane. These obstacles lie in a ``hyper-cubinder'' which is the Cartesian product of the hyperplane and a $(d-n)$-dimensional unit sphere. The number density of obstacles on this hyperplane (of area $S_n$)  is then
\[
\rho_n  = \rho S_n V_{d-n} / S_n = \rho_d V_{d-n} = \Phi \frac{V_{d-n}}{V_d}
\]
and the (dimensionless) density of obstacles in the subsystem is
\begin{equation} \begin{aligned}
\Phi_n &= \rho_n V_n \sigma^n = \Phi \frac{V_n V_{d-n}}{V_d} 
= \Phi \frac{\mathrm\Gamma(\frac{d}{2}+1)}{\mathrm\Gamma(\frac{d-n}{2}+1)\mathrm\Gamma(\frac{n}{2}+1)} \\
&= \Phi d^{\frac{n}{2}} \left[\frac{2^{\frac{-n}{2}}}{\mathrm\Gamma(1+\frac{n}{2})} + \OO(\frac{1}{d}) \right].
\end{aligned} \end{equation}
The radius of an obstacle on the hyperplane, $r_\mathrm{obs}$, corresponds to the distance of this obstacle from the hyperplane, $h$, with $h = \sqrt{1-r_\mathrm{obs}^2}$.
The probability density that a random obstacle is at distance $h$ to the hyperplane is proportional to the surface area of a $(d-n)$-dimensional shell of radius $h$ and centered at the projection of the obstacle center on the hyperplane, i.e., $P(h) \propto h^{d-n-1}$.
The radius distribution of the obstacles on $n$-dimensional hyperplane is thus
\begin{equation} \label{eq:lboundPn} \begin{aligned}
P_n(r_\mathrm{obs}) &= P_n (h) \left|\dv{h}{r_\mathrm{obs}}\right| \\
&\propto (\sqrt{1-r_\mathrm{obs}^2})^{d-n-1} \frac{r_\mathrm{obs}}{\sqrt{1-r_\mathrm{obs}^2}} \\
&= (1-r_\mathrm{obs}^2)^{\frac{d-n-2}{2}} r_\mathrm{obs}.
\end{aligned} \end{equation} 
Scaling $\hat{r} = r_\mathrm{obs} \sqrt{d} $ (hence $\hat{r} \in (0, \infty)$) in the large-$d$ limit gives
 \begin{equation}
P_n(\hat{r}) \propto \left(1-\frac{\hat{r}^2}{d}\right)^{\frac{d-n-1}{2}} \hat{r}.
\end{equation}
For $n = \OO(1)$ and $d \rightarrow \infty$, the probability density function of $\hat{r}$ is thus
\begin{equation}
P_n(\hat{r}) = \hat{r} \exp(-\frac{\hat{r}^2}{2}),
\end{equation}
which is proportional to the derivative of a Gaussian function.

To the best of our knowledge, the void percolation threshold for obstacles with such a radius distribution (denoted $\Phi_\mathrm{p,g1}(n)$) has not been investigated in finite $d$. Although we physically expect this threshold to be a nonzero finite constant for any given $n \ge 2$, this expectation cannot yet be formalized. Assuming it to hold, we then have that the $n$-dimensional subsystem is percolated if the obstacle density, after rescaling, is smaller than $\Phi_\mathrm{p,g1}(n)$, such that
\[ \Phi_n < \Phi_\mathrm{p,n} = \Phi_\mathrm{p,g1}(n) \cdot d^{\frac{n}{2}}.
\]
Invoking Eq.~\eqref{eq:lboundPn} we have that the full $d$-dimensional system is also percolated when
\begin{equation*} \begin{aligned}
\Phi d^{\frac{n}{2}} \frac{2^{\frac{-n}{2}}}{\mathrm\Gamma(1+\frac{n}{2})} &< \Phi_\mathrm{p,g1}(n) \cdot d^{\frac{n}{2}} &\Leftrightarrow \\
\Phi &< \Phi_\mathrm{p,g1}(n) 2^\frac{n}{2} \mathrm\Gamma(\frac{n}{2}+1) = \OO(1),
\end{aligned} \end{equation*}
A plausibly (although not completely) rigorous lower bound then immediately follows,
\begin{equation}
\Phi_\mathrm{p}(d) = \Omega(1).
\end{equation}

Choosing $n$ that grows with $d$ might result in a tighter lower bound, but evaluating this possibility would require solving for the dimensional scaling of $\Phi_\mathrm{p,g1}(n)$, which is seemingly a more involved problem than the original one. We thus leave this possibility for future consideration.

\section{Numerical Methods} \label{sec:methods}

In this section we detail the various numerical methods used to study the percolation threshold and the accompanying critical exponents.

\subsection{Percolation threshold detection} \label{sec:methods:phip}
One of the arguments for including void percolation in the simple percolation universality class also leads to an efficient algorithm for determining the void percolation threshold~\cite{kerstein1983equivalence}. The process entails mapping a configuration of obstacles onto a network that can then be analyzed using standard percolation criteria. The computational optimizations that enable us to consider systems up to  $d=9$ with this algorithm are described below.

More specifically, the network of edges of the Voronoi tessellation obtained from the obstacle positions is used to map the void percolation determination onto a bond percolation problem~\cite{kerstein1983equivalence}.
Each edge is then weighted by the circumscribed radius of the facet in the Delaunay triangulation that is dual to that edge. This weight thus corresponds to the minimum radius of the obstacles that can block this edge, and can be used to determine the percolation threshold. Computational implementations of Delaunay triangulation under periodic boundary conditions are, however, currently only available for $\mathbb{Z}^d$ lattices in dimensions, $d=2$ and $3$. For instance, for $d=3$, the periodic Delaunay triangulation can be built using CGAL's 3D periodic triangulation package~\cite{cgal:ct-pt3-19a}. The tessellation of the largest investigated system size ($N=10^7$), then takes several minutes per sample for a single-threaded implementation on a contemporary (Intel Xeon 6154) CPU architecture.
For $d>3$, although a comparable algorithm has been proposed~\cite{caroli2011delaunay,caroli2016delaunay}, no implementation is yet available. Even if one were, because for a fixed $N$ the number of Delaunay cells and facets grows exponentially with $d$, a full tessellation would rapidly fall beyond computational reach---limited by the available working memory---as $d$ increases.

This memory constraint is here sidestepped using two key optimizations. First, the tessellation is built locally, point-by-point~\cite{charbonneau2013geometrical,morse2014geometric}.
For each obstacle, $p_i$, we collect all nearby obstacles (and their periodic images) within a preassigned distance that guarantees all Voronoi neighbors to be included.
Then, the convex hull~\cite{Quickhull2016} of the inverse coordinates of the other obstacles is obtained after translating $p_i$ to the origin.
The vertices of this convex hull construction are then the same as the neighbors of $p_i$ in the Voronoi tessellation, but can be determined using less working memory~\cite{boissonnat2005convex}.
Second, local tessellation enables us to drop on-the-fly edges and vertices with such small weights that in a percolating network they are guaranteed to be blocked.
Visited but dropped elements can be distinguished from non-visited elements through careful bookkeeping. 
As an illustration, a $d=8$ system with $N = 10^5$ obstacles only needs $0.5\%$ of the vertices to be explicitly stored, which then take up at most 10~GB of memory.  
In total, these algorithmic improvements reduce memory usage by more than two orders of magnitude, without significantly increasing the overall computational burden. 

Once the tessellation is obtained, the percolation threshold is determined using an algorithm akin to that used for sphere percolation~\cite{newman2001fast,mertens2012continuum}. The approach uses a disjoint-set forest data structure, which efficiently organizes Voronoi vertices. For each vertex, we maintain a parent pointer and the displacement vector to its parent node. The structure thus traces back to a unique root node of the set, and each disjoint-set corresponds to a single cavity. A high-level description of the algorithm is as follows:

% To split algorithm in two pages, use following commands:
%%%%%%
% \algstore{myalg}
% \end{algorithmic}
% \end{algorithm}
% \begin{algorithm}[H]
% \begin{algorithmic} [1]
% \algrestore{myalg}
%%%%%%
\begin{algorithm}[H]
\caption{Percolation detection}
\textbf{Input:} Graph $G(X, E)$ of the Voronoi tessellation \\
\textbf{Output:} Percolating obstacle radius $\sigma_\mathrm{obs}$
\begin{algorithmic}[1]
\State Sort $E$ in descending order by weight $w_i$
\For{$e_i \in E$}
  \State Get vertices $(X_1, X_2)$ and $w_i \leftarrow e_i$
  \If{$\mathrm{root}(X_1) \neq \mathrm{root}(X_2)$}
    \State \Call{Merge}{$\mathrm{root}(X_1), \mathrm{root}(X_2)$}
  \ElsIf{\Call{CheckPercolation}{$X_1, X_2$}}
      \State $\sigma_\mathrm{obs} \leftarrow w_i$
      \State \textbf{break}
  \EndIf
\EndFor
\end{algorithmic}
\end{algorithm}

In other words, if two vertices, $X_1$ and $X_2$, do not yet belong to a same cavity, then a standard merging operation is conducted; otherwise, percolation is checked as follows:
\begin{enumerate}
\item calculate the displacement vector (under periodic condition) $\bm{r}_0 = X_1 - X_2$;
\item calculate the displacement vectors $\bm{r}_1$ and $\bm{r}_2$, from $X_1$ and $X_2$ to the root, respectively;
\item test $\bm{r}_1 - \bm{r}_2 \neq \bm{r}_0$.
\end{enumerate}
If the displacements obtained from the two methods differ (necessarily, by an integer multiple of the box side), then the cavity must form a cycle across the periodic boundary and thus percolate. If percolation is detected, then the threshold is calculated by Eq.~\eqref{eq:densitydef} setting the obstacle diameter to be the weight of last merged edge, i.e., $\sigma_\mathrm{obs} = w_i$.  

For a finite-size periodic system, different definitions of the percolation threshold have been suggested~\cite{mertens2012continuum}, including
\begin{itemize}
\item $\Phi^e_\mathrm{p}(N)$ -- there exists a percolated cavity in any coordinate;
\item $\Phi^h_\mathrm{p}(N)$ -- there exists a percolated cavity in a specific coordinate;
\item $\Phi^b_\mathrm{p}(N)$ -- there exists a percolated cavity in all $d$ coordinates.
\end{itemize}
These definitions are expected to converge to a same threshold value in the thermodynamic limit, $\Phi_\mathrm{p}=\Phi_\mathrm{p}(\infty)$. From the critical analysis, they are also all expected to asymptotically scale, albeit with different prefactors, as~\cite{stauffer1994percolation},
\begin{equation} \label{eq:finitephip}
\Phi_\mathrm{p}(N) - \Phi_\mathrm{p}(\infty) \sim N^{-1/d\nu},
\end{equation}
where the reference values for the simple percolation universality class are taken for $\nu$ (see Table~\ref{tab:threshold}).

A periodic Delaunay triangulation (as well as its dual Voronoi tessellation) is valid if the intersection of any two intersecting Delaunay cells is a simplex~\cite{caroli2011delaunay}. In our system this condition is equivalent to saying that all neighbors of an obstacle in the Voronoi tessellation are distinct, i.e. a periodic copy appears only once. The minimum valid system size $N_\mathrm{min}$ of a specific type of periodic box is then proportional to the obstacle density under which an obstacle and its nearest image is scaled with unit length. This constraint corresponds to the packing fraction of the lattice at which this periodic box lies. 

In order to curb the growing box shape anisotropy of standard cubic boxes (forming a $\mathbb{Z}^d$ lattice) as dimension increases---and thus minimize $N_{\mathrm{min}}$---we consider other $d$-dimensional (periodic boundary) simulation boxes, such as the Wigner-Seitz cell of the $D_d$ checkerboard lattices (the densest packing of spheres in $d=3$, 4 and 5) as well as the $E_8$ and $D_9^{0+}$ lattices (the densest packing of spheres in $d=8$ and $9$, respectively)~\cite{convay1982fast}.
Note that the $E_8$ and $D_9^{0+}$ lattices are special cases of the $D_n^+/D_n^{0+}$ family of lattices which for $d \ge 8$ have twice the packing fraction of $D_d$ lattices.

Uniformly distributed random obstacles are then generated as follows. 
For the unit-side $\mathbb{Z}^d$ periodic box, random vectors $X^d \in [-0.5, 0.5)^d$ are simply generated in sequence.
For both $D_d$ and $D_n^+/D_n^{0+}$ boxes, a random vector $X^d \in [-1, 1)^d$ is first generated, and the minimum-image convention with respect to the origin is then applied.
Because the simple cubic lattice (with nearest neighbor point distance of $2$) is a sublattice of both $D_d$ and $D_n^+/D_n^{0+}$, such cubic periodic box contains integer numbers of these non-cubic periodic boxes. The generated points can thus be folded back to the periodic box via the minimum image transformation, while keeping a uniform obstacle distribution.

For a given $d$, the ratio of $N_\mathrm{min}$ (as well the lattice packing fraction) gives the relative performance (RP) of a box geometry $\mathbb{P}$ compared to the conventional cubic box as $\mathrm{RP}(\mathbb{P})$. In particular, we expect $D_d$ boxes to have
\begin{equation} \label{eq:rpdn}
\mathrm{RP}(D_d) = 2^{(d-2)/2},
\end{equation}
and $D_n^+/D_n^{0+}$ boxes in $d \ge 8$  to have
\begin{equation} \label{eq:rpdnplus}
\mathrm{RP}(D_n^+/D_n^{0+}) = 2^{d/2}.
\end{equation}
The associated methodological improvement extends the length of the computationally accessible asymptotic regimes to smaller system sizes, resulting in over an order of magnitude speed up in extracting $\Phi_\mathrm{p}$ in $d=8$ and $9$. The computational complexity of the tessellation, however, appears to be super-exponential with $d$. Because evaluating a single local convex hull costs 20~s in $d=8$ and 3~min in $d=9$ on a contemporary (Intel Xeon 6154) CPU architecture, higher dimensions are thus computationally inaccessible at this time.

\subsection{Cavity reconstruction} \label{sec:methods:cavity}
In order to assess the caging criticality, we notably consider the cavity volume distribution. While infinite systems below the percolation threshold contain both an infinite volume cavity as well as large finite cavities, those above the percolation threshold contain cavities that are mostly small. Sampling them is then amenable to a cavity reconstruction scheme that is a limit case of the approach used for the Mari-Kurchan model~\cite{charbonneau2014hopping}. (Obstacles are here hard and monodisperse in size, instead of soft and size polydisperse.) This approach offers a marked computational advantage over standard simulations boxes in that it eliminates any putative sampling bias introduced by the use of periodic boundary conditions.

The overall procedure consists of placing obstacles within a finite spherical shell centered around the origin, with unit inner radius---to makes sure the origin in uncovered---and outer radius  $r_\mathrm{max}>1$, and of considering only the cavity that contains the origin. The properties of such reconstructed cavities track those of an infinite system within the shell of thickness $r_\mathrm{max}-1$.  
More specifically, like the clusters generated by the Leath algorithm for a lattice systems~\cite{leath1976clustersize}, the cavities generated by cavity reconstruction for the RLG are evenly sampled in a site base. In other words, the probability of generating a cavity of volume $V_\mathrm{cavity}$ is proportional to $V_\mathrm{cavity} P(V_\mathrm{cavity})$, where $P(V_\mathrm{cavity})$ is the probability of having a cavity of volume $V_\mathrm{cavity}$ in the thermodynamic limit. Rare large cavities that are not closed at $r_\mathrm{max}$, 
however, give rise to an undersampling bias. In order to limit this effect,  $r_\mathrm{max}$ 
is chosen such that fewer than $0.2\%$ of the cavities are not closed. The largest achievable $r_\mathrm{max}$  
nonetheless limits how close the percolation threshold can be approached with this scheme, because upon approaching $\Phi_\mathrm{p}$ increasingly large cavities dominate. 

To account for obstacle number fluctuations within finite volumes, the number of obstacles $N$ to be placed within a shell is chosen at random from the Poisson distribution
\begin{equation} \label{eq:poisson}
p(N) = \frac{N_0^N e^{-N_0}}{N!},
\end{equation}
with $N_0 =\Phi (r^d_\mathrm{max} - 1)$,
the expected number of obstacles for the system size and density considered. These $N$ obstacles are then placed uniformly at random within the hyperspherical shell, which is achieved by sequentially generating vectors of random orientation and of norm 
\begin{equation}
|r| = \left[ x(r^d_\mathrm{max} - 1) + 1 \right]^{1/d},
\end{equation}
where $x$ is a random variable uniformly distributed over $[0, 1)$.

\begin{figure}[h]
\includegraphics[width=1\columnwidth]{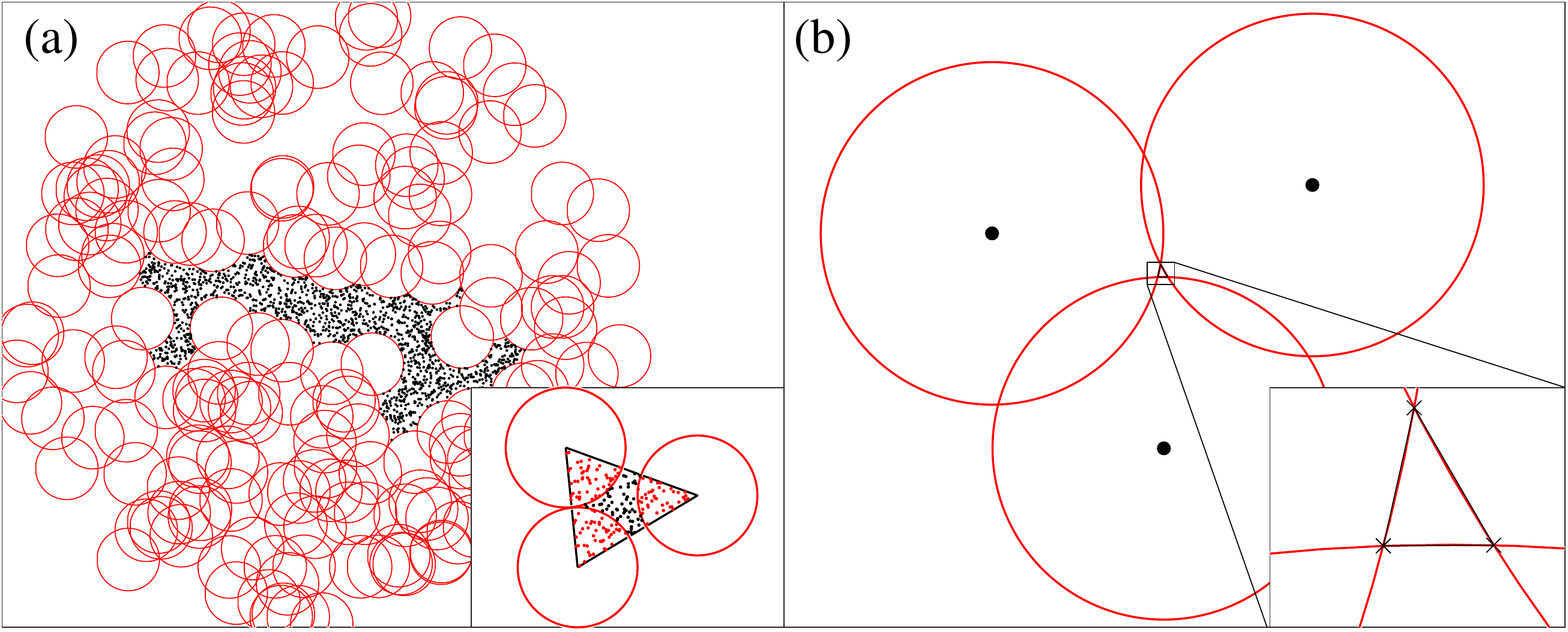}
\caption{(a) Points in $S_\mathrm{voids}$ (black dots) obtained by sampling uniformly at random a cavity closed by obstacles (red circles). (Inset) Full set of samples in a Delaunay cell, including $S_\mathrm{voids}$ (black dots) and rejected points (red dots). (b) For small cavities, the vertices of the cavity (crosses) can be used to build a second triangulation.}
\label{fig:sampling}
\end{figure}

The span of the cavity can then be obtained using an algorithm adapted from Sastry \textit{et al.}, who showed that all and only the void space that belongs to a given cavity is obtained from this approach~\cite{sastry1997statistical}. A Delaunay triangulation, which divides space into $d$-simplicial cells, is first established using CGAL's $d$D Triangulation library~\cite{cgal:hdj-t-19a}. The cavity is then constructed by running a graph search using cells as vertices and facets as edges. Starting from the cell that contains the origin, an edge (facet) is connected if the circumcenter of two cells are on same side of that facet, or if the circumcenters are on opposite sides of that facet and the facet's circumradius is greater than $\sigma_\mathrm{obs}$. All visited cells are added to the cavity. Sastry \textit{et al.} also introduced an exact algorithm for determining the cavity volume through a recursive division of $d$-simplices, but this decomposition into simple primitives is quite involved in general $d$. We thus instead use a random sampling algorithm. The idea is to generate points (samples) uniformly at random within the cavity and to use these samples to approximate the cavity volume and other physical quantities, as illustrated in Figure~\ref{fig:sampling}(a). The high-level description of the algorithm is as follows:
% To split algorithm in two pages, use following commands:
%%%%%%
% \algstore{myalg}
% \end{algorithmic}
% \end{algorithm}
% \begin{algorithm}[H]
% \begin{algorithmic} [1]
% \algrestore{myalg}
%%%%%%
\begin{algorithm}[H]
\caption{Sampling a cavity}
\begin{algorithmic}[1]
\For{$C_i\in$ visited cells}
  \State $V_i \leftarrow$ \Call{SimplexVolume}{$C_i$}
  \State Increment $V_\mathrm{cells}$
\EndFor
\For{$j=1$ to $N_\mathrm{samples}$}
  \State Randomly select simplex $C_k\in\{C_i\}$ with probability $V_k/V_\mathrm{cells}$
  \State Place sample $S \leftarrow$ \Call{SampleSimplex}{$C_k$} uniformly at random
\algstore{myalg}
\end{algorithmic}
\end{algorithm}
\begin{algorithm}[H]
\begin{algorithmic} [1]
\algrestore{myalg}
  \If{$S$ in the void space}
    \State Add $S$ to the void sample list, $S_\mathrm{voids}$
    \State Increment $N_\mathrm{voids}$
  \EndIf
\EndFor
\end{algorithmic}
\end{algorithm}
Basically, for each sample point we first choose one of the constitutive simplices with probability proportional to its volume 
\begin{equation}
V_\mathrm{simplex} = \left| \frac{1}{d!} \det(p_1-p_0, p_2-p_0, ..., p_d-p_0) \right|,
\end{equation} 
and then choose a random position within the selected simplex. Obtaining uniformly distributed samples within a $d$-simplex is equivalent to generating $d+1$ random spacings, $x_0, ..., x_d$, with unit sum~\cite[p.~568]{devroye1986nonuniform}. The latter step involves first generating $d$ independent and uniformly distributed random variables $y_1, ..., y_d\in[0, 1)^d$ and then sorting them in place. Taking $y_0 = 0$ and $y_{d+1} = 1$, one then has $x_i = y_{i+1} - y_i$. The random sample in this simplex is finally $S = \sum_{i=0}^d x_i p_i$. Determining whether $S$ is part of the void space requires a nearest-neighbor query of the obstacles. Note that although the obstacle nearest to $S$ is most likely one of the vertices of $C_i$, outliers are possible. This determination is accelerated by pre-computing the point-to-simplex distances for obstacles other than the simplex vertices, and storing obstacles with distance less than $\sigma_\mathrm{obs}$ as candidate nearest neighbors.

As the obstacle density increases, the fraction and size of the voids become increasingly small. Because the probability of a sample lying in a void follows the binomial distribution, the variance for the number of voids $N_\mathrm{voids}$ (out of $N_\mathrm{samples}$ samples) is $\mathrm{var}(N_\mathrm{voids}) \approx N_\mathrm{voids}(1 - N_\mathrm{voids}/N_\mathrm{samples})$, and the sampling error then also grows large. For sufficiently small cavities, we consider an alternate sampling scheme that sidesteps this difficulty. As illustrated in Figure~\ref{fig:sampling}(b), the approach consists of identifying the vertices of this cavity, building a triangulation over them, and then running the cavity sampling algorithm for the new triangulation.  The fraction of void samples ($N_\mathrm{voids}/N_\mathrm{samples}$) then markedly increases, which reduces the sampling error.
Because a simplex generated this way may lie completely in occupied space or even contain the voids of other cavities, however, a certain caution must be exercised. Here, it is only invoked if the original sampling first failed to find fewer than $1000$ void samples out $N_\mathrm{samples} = 2\times10^5$, which corresponds to a relative error of about $\sim 3\%$ per cavity in the original sampling scheme. In practice, this stringent criterion suffices to completely prevent geometrical complications.

Given samples inside the void space, $S_i\ (i \in [1, N_\mathrm{voids}])$, different observables (with length still given in units of $\sigma$) can be computed:
\begin{itemize}

\item the volume
\begin{equation}
V_\mathrm{cavity} = V_\mathrm{cells} \frac{N_\mathrm{voids}}{N_\mathrm{samples}},
\end{equation}
where $V_\mathrm{cells}$ is the total volume of the cells considered;

\item the infinite-time MSD of a tracer, i.e., the cage size, 
\begin{equation}
\Delta_\mathrm{cavity} = \langle (S_i - S_j)^2 \rangle = 2(\langle S_i^2 \rangle - \langle S_i \rangle^2 );
\end{equation}

\item the long-time limit of the self van Hove function, $G_\mathrm{s}(r, t)$, which is the probability of finding a tracer having displaced by $r$ after a time $t$,
\begin{equation} \begin{aligned}
G_\mathrm{s,cavity}(r) &= G_\mathrm{s,cavity}(r, t \rightarrow \infty) \\
&\sim \sum_{i \neq j} \delta(|S_i - S_j|-r),
\end{aligned} \end{equation}
\end{itemize}
This last result follows from every cavity site being equally probable---independent from the (random) initial position---in that limit.
The summation over sites $i \neq j$ then eliminates the artificial discretization peak at $r = 0$. Note that the expected $V_\mathrm{cavity}$, $\Delta$ and $G_\mathrm{s}(r)$ are obtained by taking the arithmetic mean over all randomly generated cavities.

\subsection{Dynamics} \label{subsec:dyn}
The tracer dynamics is obtained from a high-dimensional generalization of the simulation scheme described by H{\"o}fling \emph{et al.}~\cite{hofling2006localization,hofling2008critical}. Specifically, $N$ obstacles are placed uniformly at random within a $d$-dimensional periodic box, where $N=10^5-10^6$, with the upper system size limit only being used for $\Phi$ close to $\Phi_\mathrm{p}$ ($6 \le \Phi \le 7$ in $d=4$ and $8.5 \le \Phi \le 9.5$ in $d=5$). 
(Unlike for the cavity reconstruction scheme, $N$ is here kept fixed, because relative size fluctuations under a Poisson field scale as $o(N^{-1/2})$, and are thus negligible.) 
A tracer is then placed at the origin and assigned an initial velocity $\dot{r}= 1$  %$d r/d t = \sigma$ 
with random orientation, and event-driven molecular dynamics then identifies the elastic collisions of the tracer with the obstacles, until the simulation ends at time $t_\mathrm{max}$.

To accelerate simulations, obstacle neighbor lists are used. (So are cell lists when system sizes warrant it, but this only happens in $d\le 4$.)
Because the computational performance in high $d$ depends sensitively on choice of cutoff radius for the neighbor list, $r_\mathrm{cut}$ and that the optimal $r_\mathrm{cut}$ depends strongly on cavity geometry, for $d \ge 5$ dynamical adjustments are made to $r_\mathrm{cut}$, such that neighbor list updates occupy $20\% -50\%$ of the overall simulation time.
In order to average over thermal noise, the MSD of a given sample at each sampled time (except $t_\mathrm{max}$) is averaged over $2^{10}$ initial times; the MSD is further averaged over at least $100$ independent replicates (and up to $200$ for the quantitative determination of critical exponents).

\begin{figure}[h]
\includegraphics[width=1\columnwidth]{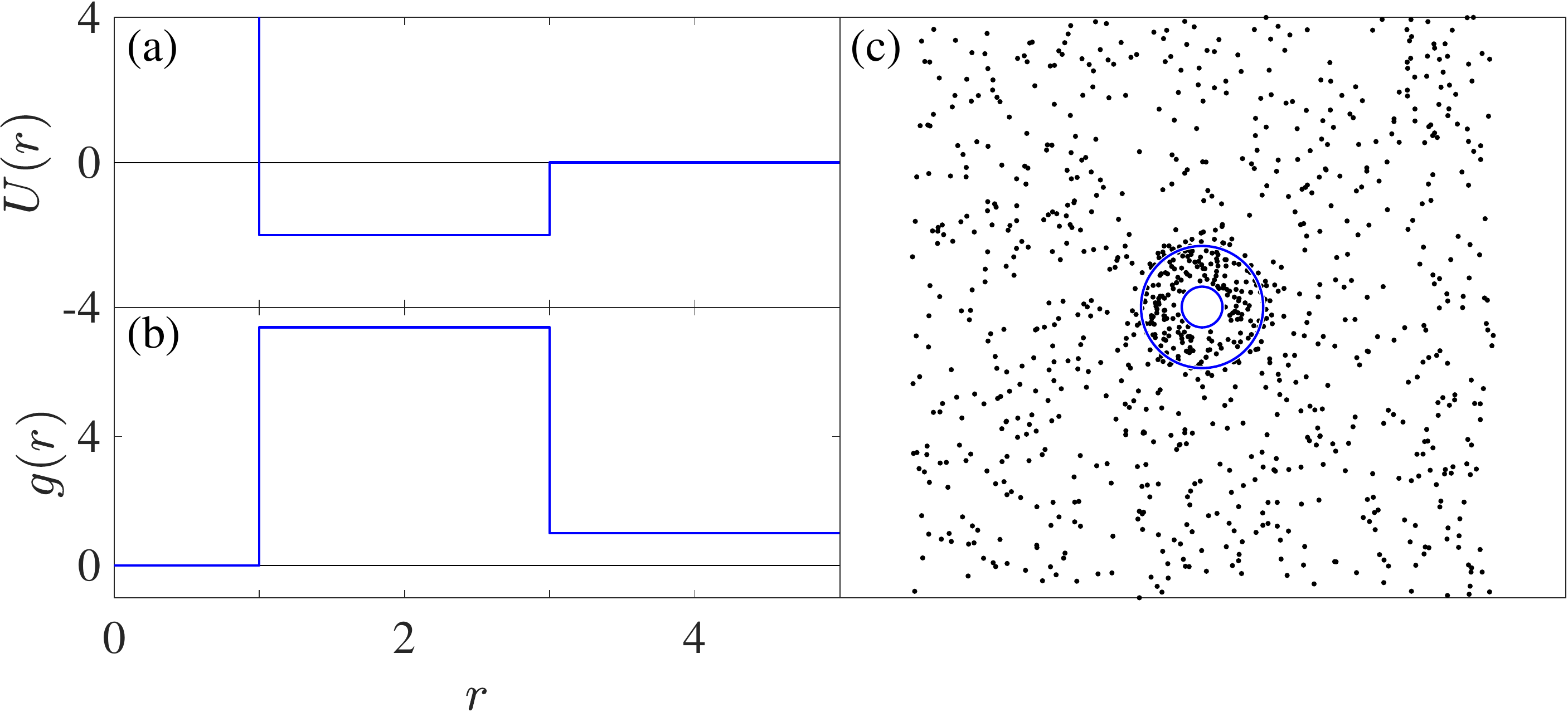}
\caption{(a) Square-well potential, $U(r)$, with $\xi=2, \lambda=2$, and (b) the resulting radial distribution function $g(r)=e^{-U(r)}$ of obstacles around the origin. (c) Sample obstacle centers for $\Phi=4$ in  $d=2$; obstacles within the square-well act as an inner shell of higher density, $\Phi' = \Phi e^{\xi}$.}
\label{fig:sqwell}
\end{figure}

We also investigate the effect of the local obstacle configuration on the tracer dynamics. Seen from the origin, the random Poisson field that controls the distribution of obstacles is akin to the potential field of a hard-sphere of radius $\sigma$. To increase caging, we include an attractive square-well potential of relative thermal strength $\xi$ and range $\lambda$. The local density of the obstacles then becomes 
\begin{equation} \label{eq:sqwell}
\Phi'(r) = \begin{cases}
0, & r < 1, \\
\Phi e^{\xi}, & 1 \le r < 1 + \lambda, \\
\Phi, & r \ge 1 + \lambda,
\end{cases}
\end{equation}
as illustrated in Fig.~\ref{fig:sqwell}.

The tracer dynamics is then run as above, but because the system is no longer translationally invariant, averaging over thermal disorder is no longer possible within a single trajectory. To compensate, at least $10^4$ replicates are used to improve the averaging.  

\section{Results and discussion} \label{sec:results}

\begin{figure*}[t]
\includegraphics[width=1\textwidth]{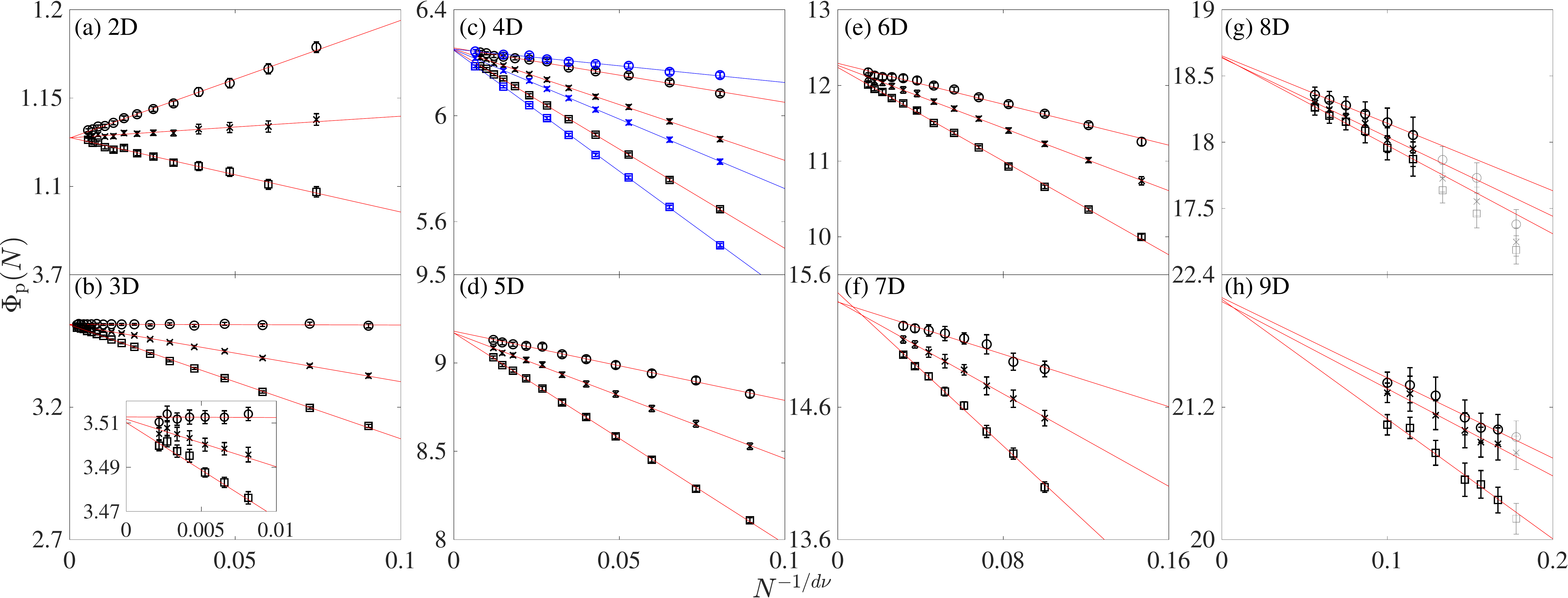}
\caption{(a-h) Finite-size scaling of $\Phi^e_\mathrm{p}$ (circles), $\Phi^h_\mathrm{p}$ (crosses) and $\Phi^b_\mathrm{p}$ (squares) in $d=2$ to $9$, with error bars denoting $95\%$ confidence intervals. All three estimates coincide, within numerical uncertainty, in the $N \rightarrow \infty$ limit. Gray markers denote small systems excluded from the fit to Eq.~\eqref{eq:finitephip}. Cubic periodic boxes are used in $d=2$ and $3$, $D_d$ periodic boxes are used in $d=4$ to $7$, and $E_8$ and $D_9^{0+}$ boxes in $d=8, 9$, respectively. In panel (c) results for $\mathbb{Z}^4$ periodic boxes (blue) are reported for comparison. As expected, the scaling form and the intercept obtained in $\mathbb{Z}^4$ and $D_4$ boxes are consistent, and finite-size corrections of the latter are significantly smaller. The inset in (b) enlarges the large $N$ regime in $d=3$, in order to precisely determine $\Phi_\mathrm{p} = 3.510(2)$.
}
\label{fig:threshold}
\end{figure*}

In this section, we evaluate the percolation threshold and critical properties of the RLG and void percolation by the numerical methods described in Sec.~\ref{sec:methods} and compare the results with theoretical predictions described in Sec.~\ref{sec:scaling} and~\ref{sec:bounds}.

\subsection{Percolation threshold}

Our percolation threshold detection algorithm increases the upper range of accessible system size by orders of magnitude compared to earlier works, which makes the asymptotic power-law fitting regime to Eq.~\eqref{eq:finitephip} fairly robust in all dimensions considered  (Fig.~\ref{fig:threshold}). As further validation, we make sure that the percolation thresholds evaluated through the three criteria described in Section~\ref{sec:methods:phip} all coincide in the thermodynamic $N \rightarrow \infty$ limit. However, only the extrapolation results of $\Phi^b_\mathrm{p}(N)$, which offers the smallest variance and the widest asymptotic regime~\cite{newman2000efficient,jin2015dimensional}, are used to estimate $\Phi_\mathrm{p}$ (Table~\ref{tab:threshold}). 

\begin{table}[ht]
\caption{Numerical estimates of the void percolation threshold}
\begin{tabular}{cccc}
\hline \hline
$d$ & $\Phi_\mathrm{p}$ (this work) & Other sources  \\
\hline
2 & 1.1276(9) & 1.12808737(6)~\cite{mertens2012continuum}, 1.121(2)~\cite{jin2015dimensional} \\
3 & 3.510(2) & 3.506(8)~\cite{hofling2008critical}, 3.515(6)~\cite{yi2012computational},\\ 
& & 3.500(6)~\cite{jin2015dimensional} \\
4 & 6.248(2) & 6.16(1)~\cite{jin2015dimensional} \\
5 & 9.170(8) & 8.98(4)~\cite{jin2015dimensional} \\
6 & 12.24(2) & 11.74(8)~\cite{jin2015dimensional} \\
7 &  15.46(5) & - \\
8 & 18.64(8) & - \\
9 & 22.1(4) & - \\
\hline
\end{tabular}
 \label{tab:threshold}
\end{table}

The numerical estimates for the percolation threshold are generally consistent with previously reported results (Table~\ref{tab:threshold}). As described in Sec.~\ref{sec:bounds}, in $d=2$ the percolation thresholds of the spheres and their void spaces are provably the same, hence an algorithmic route much more efficient than the Voronoi construction can be used~\cite{mertens2012continuum}. Our direct result is consistent with that estimate, albeit orders of magnitude less accurate.
In $d=3$ and above, the two percolation thresholds are no longer identical. In fact, void percolation thresholds are significantly larger than those from direct percolation, which further hampers their computation. Void percolation thresholds can nonetheless be determined up to a few parts in a hundred up to $d=9$. In $d=3$, our results, while still consistent with published results, are the most accurate of the lot. For $4 \le d \le 6$, our results are systematically larger than the only reported values~\cite{jin2015dimensional}, a discrepancy likely arising from that earlier effort having included systems with $N<N_\mathrm{min}$ in fitting the asymptotic scaling form. 

\begin{figure}
\includegraphics[width=1\columnwidth]{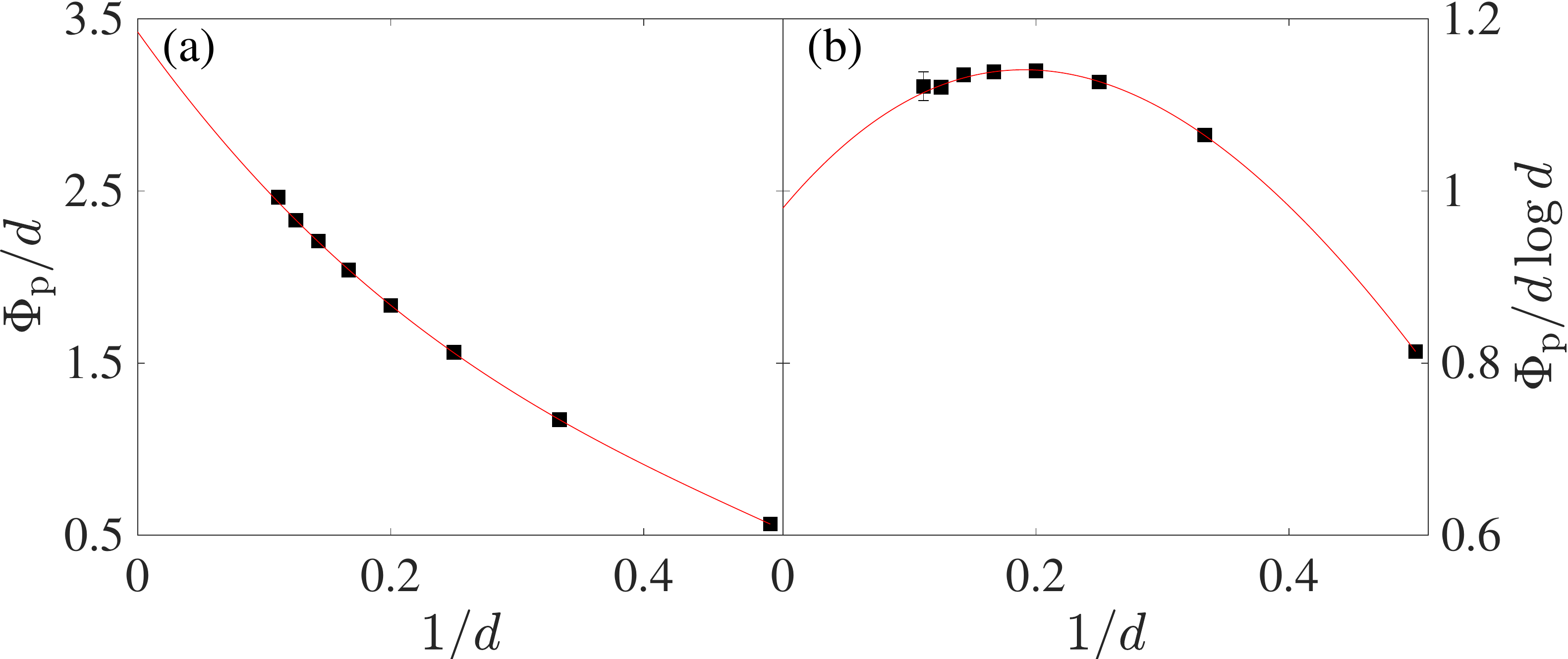}
\caption{Dimensional scaling of the percolation threshold of $d=2$ to $9$ (squares, from right to left) with the proposed scaling form (a) $\Phi_\mathrm{p} \sim d$ and (b) $\Phi_\mathrm{p} \sim d \log d$. Red lines denote the cubic polynomial fitting. In (a) the prefactor of the scaling grows smoothly and monotonically with dimension; in (b), however, the dimensional dependence of the prefactor is non-monotonic.}
\label{fig:phipscale}
\end{figure}

Because results smoothly evolve across dimension, and because no finite dimension above $d_\mathrm{u}$ is expected to exhibit singularly new physics, we attempt to extrapolate our results to higher $d$. In the context of the mean-field theory for the RLG~\cite{biroli2020unifying}, a natural asymptotic $d\rightarrow\infty$ scaling is $\Phi_\mathrm{p}(d) \sim d$. Fitting the results  (excluding $d = 9$ which has a relatively large uncertainty) to a cubic form of this type then nicely gives (Fig.~\ref{fig:phipscale}(a))
\begin{equation}
\label{eq:scalefit}
\Phi_\mathrm{p} = d \left[3.42(8) - 10.3(9) /d + 13(3) /d^2  -9(4) /d^3\right],
\end{equation}
with $\Phi_\mathrm{p}(d \rightarrow \infty) = d \cdot 3.40(5)$. Note that because $\Phi_\mathrm{p}/d$ grows convexly with $1/d \rightarrow 0$, this form provides a lower bound to the actual asymptotic $d\rightarrow\infty$ result.
An alternate possibility would be to use a scaling that saturates the upper bound from in Sec.~\ref{sec:phipup}, $\Phi_\mathrm{p} \sim d \log d$. Fitting the results to this form with a cubic correction then gives
\begin{equation}
\label{eq:scalefit2}
\Phi_\mathrm{p} = d \ln d \left[0.98(5) + 1.8(6)/d - 5.2(21) /d^2+ 2(2) /d^3\right].
\end{equation}
As shown in Fig.~\ref{fig:phipscale}(b), the prefactor to this second scaling decreases for $d > 5$, suggesting that this proposed form is indeed an (possibly saturated) upper bound to the actual asymptotic $d\rightarrow\infty$ result. The lower and upper bounds suggested from numerical simulation, $3.60 d \le \Phi_\mathrm{p} \le d \log d$ thus lie between the analytical bounds of Section~\ref{sec:bounds}. (In all cases, they grow more than exponentially faster than direct sphere percolation threshold, $\Phi_\mathrm{c} \sim 2^{-d}$~\cite{torquato2012effect1}.)
These findings motivate fully formalizing these bounds as well as attempting to tighten them further.

\subsection{Structural percolation exponents}

\begin{table*}[t]
\caption{Geometric critical exponents in void and lattice percolation}
\begin{tabular}{cccccccc}
\hline \hline
Dimension & $\nu$~\cite{koza2016discrete} & $\tau$ & (lattice)~\cite{mertens2018percolation} &  $\gamma$ & (lattice)~\cite{stauffer1994percolation} & $\mu_-$ & (lattice)~\cite{biroli2019dynamics}  \\
\hline
2 & 4/3~\cite{stauffer1994percolation} & 2.0549(6) & $187/91 \approx 2.0549$~\cite{stauffer1994percolation} & 2.38(2) & $43/18\approx2.389$ &  2.50(3) &  $91/36 \approx 2.528$~\cite{stauffer1994percolation} \\
3 & 0.8774(13) & 2.179(1) & 2.1892(1) & 1.76(4) &  1.80 & 1.35(5) & 1.3377(15)\\ 
4 & 0.6852(28) & 2.295(2) & 2.3142(5) & - & 1.44 & - & 0.73(1) \\
5 & 0.5723(18) & - & 2.419(1) & - & 1.18 & - & 0.31(2) \\
$\ge 6$ & 1/2 & - & 5/2 & - & 1 & - & 0 (logarithmic) \\
\hline
\end{tabular}
 \label{tab:structexp}
\end{table*}

The cavity reconstruction scheme described in Section~\ref{sec:methods:cavity}  is used to examine the criticality of the percolating cluster at $\Phi_\mathrm{p}$ as well as the growth of the mean cluster size upon approaching $\Phi_\mathrm{p}$ from above. Because void percolation is part of the simple percolation universality class, its critical exponents associated with structure are expected to match those of lattice percolation. Dynamical results roughly support this expectation for $\nu$ and $\mu_-$ in $d=3$~\cite{hofling2008critical} and $\mu$ in $d=2,3$~\cite{hofling2008critical,bauer2010localization}, but few other exponents have been considered, and in $d \ge 4$ none have been considered. We here more carefully evaluate some of the geometric exponents without explicitly resorting to dynamics.

\begin{figure}
\includegraphics[width=1\columnwidth]{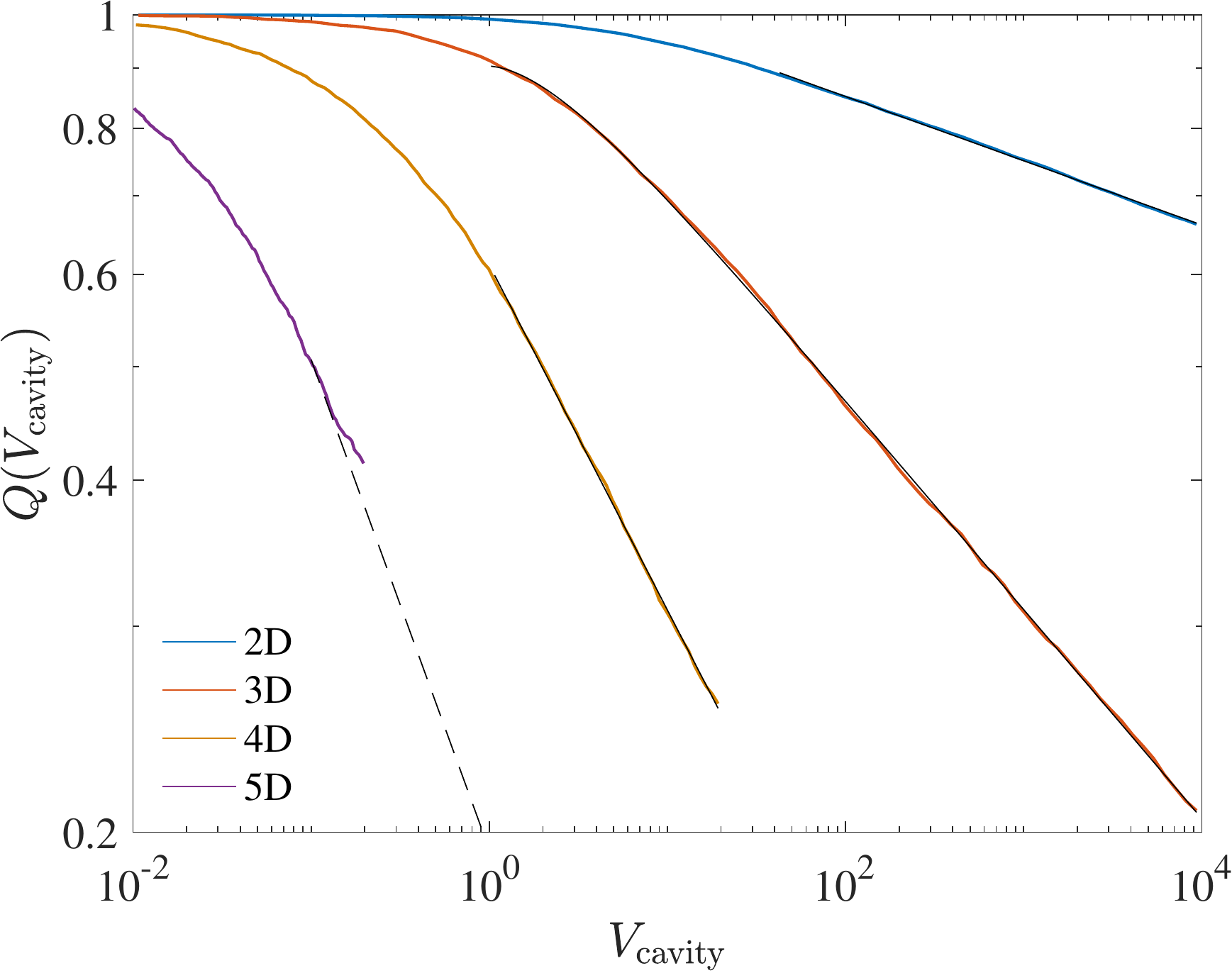}
\caption{The complementary cumulative cavity volume distribution $Q(V_\mathrm{cavity})$ at $\Phi_\mathrm{p}$ exhibits the Fisher power-law scaling. $V_\mathrm{cavity}$ is in units of $\sigma^d$. Black solid lines are fits to Eq.~\ref{eq:fisherfit} for $d=2$ and $3$. A simple linear regression is used for $d=4$ because the nonlinear fit is unstable for this limited a range. In $d=5$ a quantitative regression is not possible, hence the expected scaling is only given as reference (dashed line, see text).}
\label{fig:fisher}
\end{figure}

Recall that the cavity volume distribution at $\Phi_\mathrm{p}$ is expected to scale as $P(V_\mathrm{cavity}) \sim V_\mathrm{cavity}^{-\tau}$ for large cavities (Eq.~\eqref{eq:fisherscaling}), and that the cavity reconstruction scheme generates cavities with a probability proportional to $V_\mathrm{cavity} P(V_\mathrm{cavity})$. We can thus extract the Fisher exponent by reconstructing cavities at $\Phi_\mathrm{p}$. Numerically, it is convenient to evaluate the complementary cumulative cavity volume distribution
\begin{equation}
Q(V_\mathrm{cavity}) = \int_{V_\mathrm{cavity}}^{\infty} V' P(V') \dd V',
\end{equation}
with open cavities taken as having an infinite volume. This function is then expected to scale as 
\begin{equation*}
Q(V_\mathrm{cavity}) = C_V V_\mathrm{cavity}^{2-\tau}(1 + F_{V} V_\mathrm{cavity}^{-\Omega}),
\end{equation*}
with asymptotic and leading pre-asymptotic contributions~\cite{mertens2018percolation}, and $C_V, F_V, \Omega$ as fit parameters. The logarithmic form~\cite{mertens2018percolation}
\begin{equation} \label{eq:fisherfit}
\begin{aligned}
&\ln Q(V_\mathrm{cavity}) = \\
&(2-\tau) \ln V_\mathrm{cavity} + \ln(1 + F_{V} V_\mathrm{cavity}^{-\Omega}) + \ln C_V.
\end{aligned}
\end{equation}
is plotted Fig.~\ref{fig:fisher} and the fitted values of the critical exponents are given in Table~\ref{tab:structexp}. The extracted $\tau$ for $d=2$ is fully consistent with the exact value from lattice models, $\tau=187/91$. In $d=3, 4$, the extracted exponents are also very close to the most accurate exponents extracted from lattice models~\cite{mertens2018percolation}. Although the $1\%$ deviation lies outside the error range, the difference likely reflects our inclusion of pre-asymptotic points within the limited available fitting regime. 
In $d=5$ and beyond, however, quantitative extraction of the Fisher exponent is not possible because cavities sufficiently large to even approach the critical regime lie beyond computational reach.

\begin{figure}[ht]
\includegraphics[width=1\columnwidth]{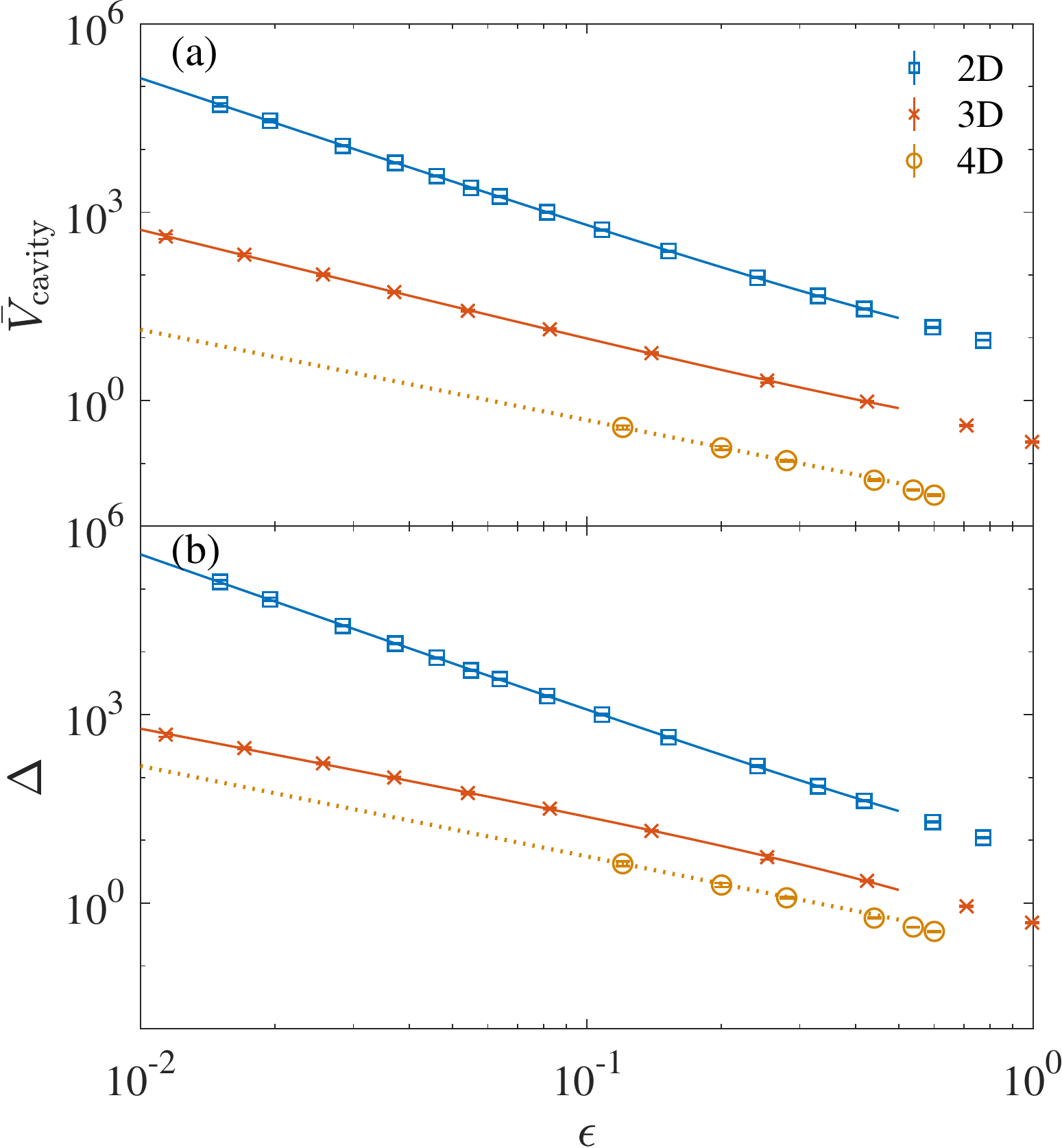}
\caption{Power-law divergence of (a) the mean cavity volume and (b) the cage size upon approaching the percolation threshold. In $d=2$ and $3$ results for $\epsilon < 0.5$ are fitted with Eq.~\ref{eq:fisherfit} (solid lines). In $d=4$ the power-law scaling using lattice exponents is provided as reference (dotted lines).
}
\label{fig:structexp}
\end{figure}

Upon approaching $\Phi_\mathrm{p}$ from above, the mean cavity volume $\bar{V}_\mathrm{cavity} = \int_0^\infty V'^2 P(V') \dd V'$ is expected to diverge as $\bar{V}_\mathrm{cavity}(\epsilon) \sim \epsilon^{-\gamma}$, where $\gamma$ is the mean cluster size exponent.
In addition, Eq.~\eqref{eq:localscaling} can be used to evaluate $\mu_-$ from $\Delta$. We thus implement variants of Eq.~\eqref{eq:fisherfit}  as fitting forms
\begin{align}
\ln \bar{V}_\mathrm{cavity} &= -\gamma \log \epsilon + F_{\bar{V}} \epsilon + C_{\bar{V}} \label{eq:fitstructgamma}\\
\ln \Delta &= -\mu_- \ln \epsilon + F_{\Delta} \epsilon + C_{\Delta} \label{eq:fitstructmu}
\end{align}
where we have approximated $ \ln(1 + F \epsilon^\Omega ) \approx F \epsilon$ with $-\Omega=1$ compared to Eq.~\eqref{eq:fisherfit}. Eliminating the logarithmic operation and higher-order corrections in the fit stabilizes the regression procedure over the $\sim 10$ available data points, which range over one and a half decade. The resulting form is plotted in Fig.~\ref{fig:structexp} and the extracted $\gamma, \mu_-$ are reported in Table~\ref{tab:structexp}.

For both $d=2$ and $3$, the extracted critical exponents are fully consistent with their lattice percolation counterpart. For $d=4$, although the regression window available presents some numerical challenges, the results are consistent with the expected scaling from the lattice exponent values. In short, our analysis strongly supports the universality of the geometric critical exponent, including $\tau, \gamma$ and $\mu_-$, for both lattice and void percolation.

\subsection{Transport percolation exponents}

\begin{table*}[t]
\caption{Transport critical exponents in void and lattice percolation}
\begin{tabular}{ccccccc}
\hline \hline
Dimension & $\mu$ & Eq.~\eqref{eq:diffusionscale} & $\mu_\mathrm{lattice}$~\cite{stauffer1994percolation} &  $d'_\mathrm{w}$ & Eq.~\eqref{eq:subdiffscale} & $d'_\mathrm{w,lattice}$~\cite{biroli2019dynamics}  \\
\hline
2 & 1.31~\cite{bauer2010localization}\footnote{\label{note:2dc}Reference~\onlinecite{bauer2010localization} did not extract critical exponents from its data, hence the precision estimates are not given.} & $(=\mu_\mathrm{lattice})$  &  1.310(1)~\cite{grassberger1999conductivity} & 3.04~\cite{bauer2010localization}\footnoteref{note:2dc} & ($=d'_\mathrm{w,lattice}$) &  3.036~\cite{grassberger1999conductivity,bauer2010localization}  \\
3 &  2.88~\cite{hofling2008critical} & 2.877(1) & 2.0 & 6.25~\cite{hofling2008critical} & 6.30(1)  & 4.94(1) \\
4 &  4.28(8) & 4.370(6) & 2.4 &  12.5(8) &  14(1) &  8.64(4) \\
5 &  5.6(2) &  5.717(5) & 2.7 &  - & $\approx39$ & 20(3) \\
$\ge 6$ &  - &  $d+1$ & 3 &  - &  \multicolumn{2}{c}{$\infty$ (logarithmic)} \\
\hline
\end{tabular}
 \label{tab:transportexp}
\end{table*}

\begin{figure}
\includegraphics[width=1\columnwidth]{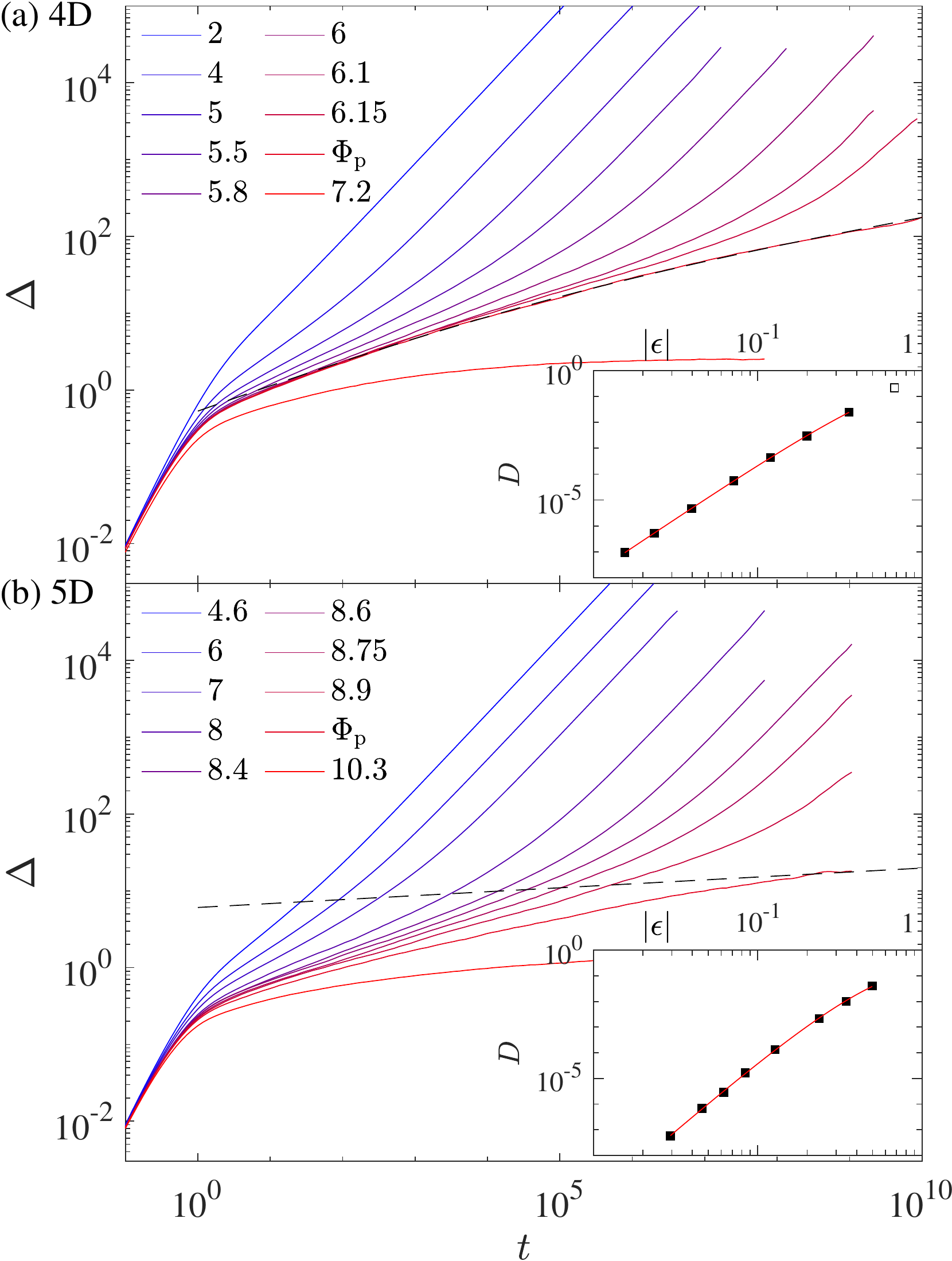}
\caption{MSD in (a) $d=4$ and (b) $d=5$ under various obstacle densities $\Phi$. For (a), fitting the subdiffusion scaling form, Eq.~\eqref{eq:subdiffusefit} (dashed line), at the threshold gives $d'_\mathrm{w} = 12.5(8)$. For (b), the dashed line denotes the expected long-time subdiffusion scaling $d'_\mathrm{w} \approx 39$ which is not yet met. (inset) The diffusion coefficients  (markers) in both (a) and (b) vanishes as a power law $D \sim |\epsilon|^\mu$ upon approaching $\Phi_\mathrm{p}$. Fitting diffusion constants with $|\epsilon|<0.5$ using the critical form, Eq.~\eqref{eq:mufit} gives  $\mu(d=4)=4.28(8)$ and $\mu(d=5)=5.6(2)$.}
\label{fig:dyn}
\end{figure}
With accurate percolation thresholds in hand, we can also evaluate transport exponents in $d > 3$, and compare the results with the theoretical prediction from Eq.~\eqref{eq:diffusionscale} and with lattice simulation results. Simulation results in $d=4$ and  $5$ are shown in Fig.~\ref{fig:dyn}.

We again adapt the power-law scaling form with sub-leading correction of Eq.~\eqref{eq:fisherfit} to analyze the subdiffusion at $\Phi_\mathrm{p}$,
\begin{equation} \label{eq:subdiffusefit}
\ln \Delta(t) = \frac{2}{d'_\mathrm{w}} t + \ln(1 + F_{t} t^{-\Omega_t}) + \ln C_t.
\end{equation}
Fitting the time range $t > 10^3$ gives $d'_\mathrm{w}(d=4) = 12.5(8)$, which is consistent with the prediction in Table~\ref{tab:transportexp}, and clearly distinct from the corresponding lattice exponent. In $d=5$, however, the associated subdiffusion exponent, $d'_\mathrm{w}(d=5) \approx 39$, leads to a fairly flat curve, making this exponent too numerically challenging to evaluate quantitatively.

For the diffusion criticality, we extract the diffusion constant $D = \lim_{t \rightarrow \infty} \dd \Delta/\dd t$ by linearly fitting the long-time results, and then adapting the fitting form of Eq.~\eqref{eq:fitstructgamma} as
\begin{equation} \label{eq:mufit}
\ln D(\epsilon) =  \mu \ln |\epsilon| + F_D t + C_\epsilon.
\end{equation}
Fitting the range $|\epsilon| < 0.5$ gives $\mu(d=4) = 4.28(8)$, and $\mu(d=5)=5.6(2)$, fully consistent with the theoretical prediction in Table~\ref{tab:transportexp}. 

\begin{figure}
\includegraphics[width=1\columnwidth]{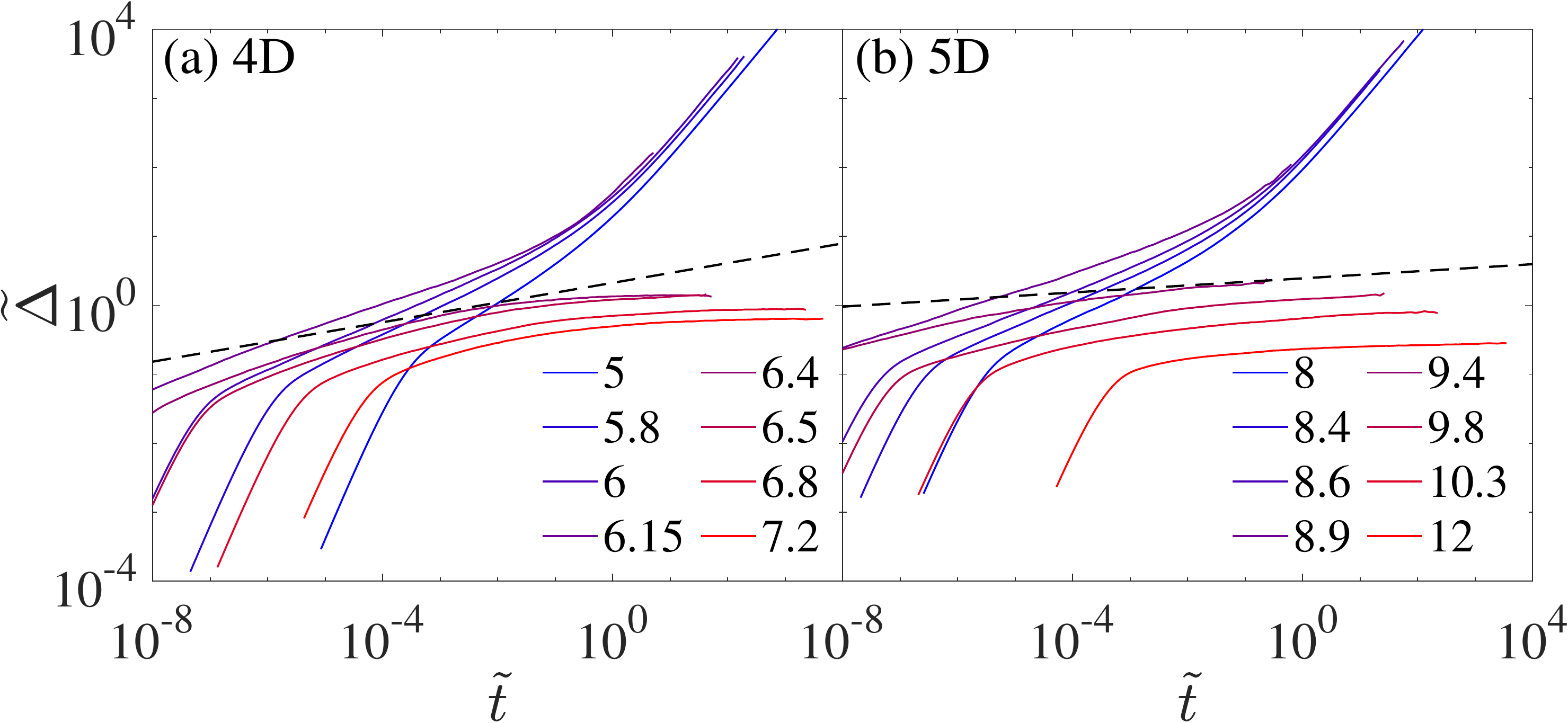}
\caption{Rescaled time and MSD for (a) $d=4$ and (b) $d=5$ in various obstacle densities $\Phi$. The dashed line denotes the subdiffusive scaling of $\tilde{\Delta} \sim \tilde{t}^{2/d'_\mathrm{w}}$ at $\Phi_\mathrm{p}=6.248$ and $9.170$ in $d=4,5$, respectively (Table~\ref{tab:threshold}).}
\label{fig:dyncollapse}
\end{figure}

Knowing the expected localization and diffusion scaling, the MSD around $\Phi_\mathrm{p}$ is expected to follow the collapse form of Eq.~\eqref{eq:uniscale}. Specifically, time and MSD are rescaled with
\begin{align*}
\tilde{t} = |\epsilon|^{(\mu_- + \mu)} t, \\
\tilde{\Delta} = |\epsilon|^\mu \Delta,
\end{align*}
which for $d=4$ and $5$ are given in Fig.~\ref{fig:dyncollapse}. In $d=4$, collapses are observed in both the diffusion and the localization regimes upon approaching the percolation threshold. In $d=5$, although the asymptotic regime is not yet fully reached, the results nonetheless suggest a collapse upon approaching the threshold from the diffusion side. On the localization side, however, the asymptotic regime is not yet reached. Such delayed onset of the asymptotic scaling regime is a signature of approaching $d_\mathrm{u}=6$, as can also be observed in lattice models~\cite{biroli2019dynamics}.

\begin{figure}[ht]
\includegraphics[width=1\columnwidth]{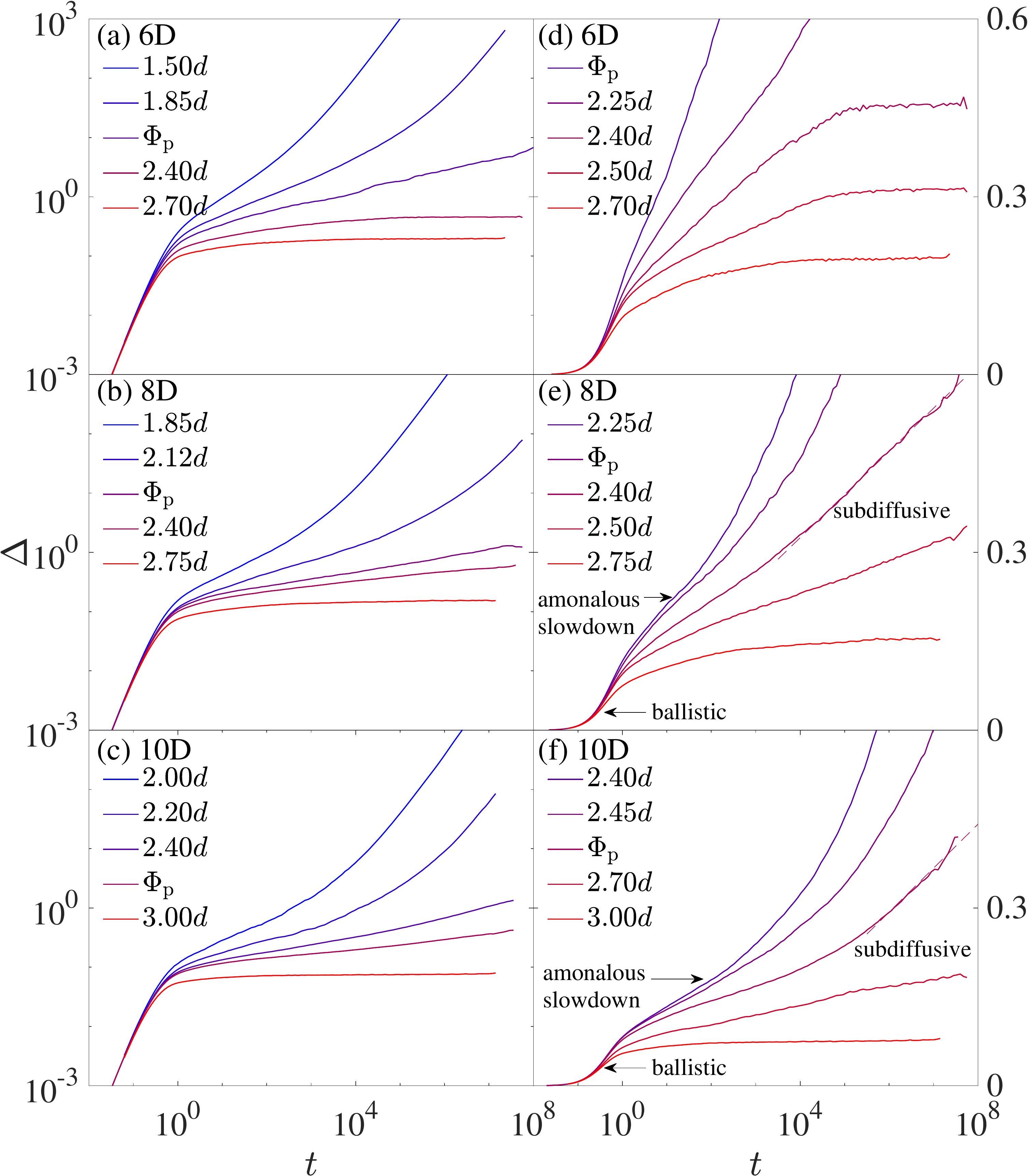}
\caption{Tracer dynamics above the upper critical dimension for various obstacle densities $\Phi$ for $d=6$, $8$, and $10$ (a-c) on a log-log scale and (d-f) on a lin-log scale. Arrows in (e, f) highlight the small-time ballistic growth and intermediate-time dynamical slowdown. Dashed lines indicate logarithmic subdiffusion around $\Phi_\mathrm{p}$. Note that to facilitate comparison between dimensions $\Phi$ is scaled by $d$ in the legend. Note also that $\Phi_\mathrm{p}$ for $d=6$ and $8$ are taken from Table~\ref{tab:threshold}, while for $d=10$ the value is extrapolated from Eq.~\eqref{eq:scalefit} to be $\Phi_\mathrm{p}=25.1(2)$. (Extrapolating using Eq.~\eqref{eq:scalefit2} gives $\Phi_\mathrm{p}=25.4(2)$.)
}
\label{fig:highddyn}
\end{figure}

Above $d_\mathrm{u}$, the power-law scaling of both localization and subdiffusion is expected from percolation theory to be replaced by a logarithmic growth. As dimension increases, the asymptotic regime is generally observed to widen as pre-asymptotic corrections diminish in range also in this direction~\cite{mertens2018percolation,biroli2019dynamics}. In lattice systems, this effect even partly cancels the computational challenge of studying higher-dimensional systems, and has made possible a clear quantitative examination of the logarithmic scaling up to $d=13$~\cite{biroli2019dynamics}. In the RLG, however, the dynamics intermediate between the short-time ballistic growth $\Delta = t^2$ and the long-time diffusion or localization regimes, grows more rather than less complex as dimension increases above $d_\mathrm{u}$ (Fig.~\ref{fig:highddyn}). In particular, the expected logarithmic subdiffusion regime is not observed at $\Phi_\mathrm{p}$ over the computationally accessible time range. As dimension increases, an intermediate dynamical slowdown clearly develops (see arrows in Fig.~\ref{fig:highddyn}(e, f)). A pre-asymptotic logarithmic growth is also clearly observed in $\Phi > \Phi_\mathrm{p}$ at intermediate times. For instance, for $2.4d \le \Phi \le 2.5 d$, the logarithmic growth of MSD survives over four decades in $d=6$ and over more than seven decades in $d=8$ (and similarly in $d=10$, although at higher $\Phi$ because the percolation threshold grows larger). 

In Ref.~\onlinecite{biroli2020unifying}, these features, which have no equivalent in lattice models, were interpreted as a finite-dimensional echo of the mean-field dynamical caging transition around $\Phi_\mathrm{d} = d \cdot 2.40339...$, which is near $\Phi_\mathrm{p}$ in this dimensional range. 
As a result, the onset of the percolation criticality is pushed beyond the computationally accessible regime. The difference between void and lattice percolation suggests that new physics emerges from the interplay of mean-field caging and percolation at these intermediate times. Further discussing the details of the mean-field description of the RLG can be found in~\cite{biroli2020mean}. Here, we instead consider a scheme for enhancing the interplay between caging and percolation by slightly modifying the RLG in Sec.~\ref{sec:inhomo}. 

\section{Inhomogeneous RLG} \label{sec:inhomo}

\begin{figure}
\includegraphics[width=1\columnwidth]{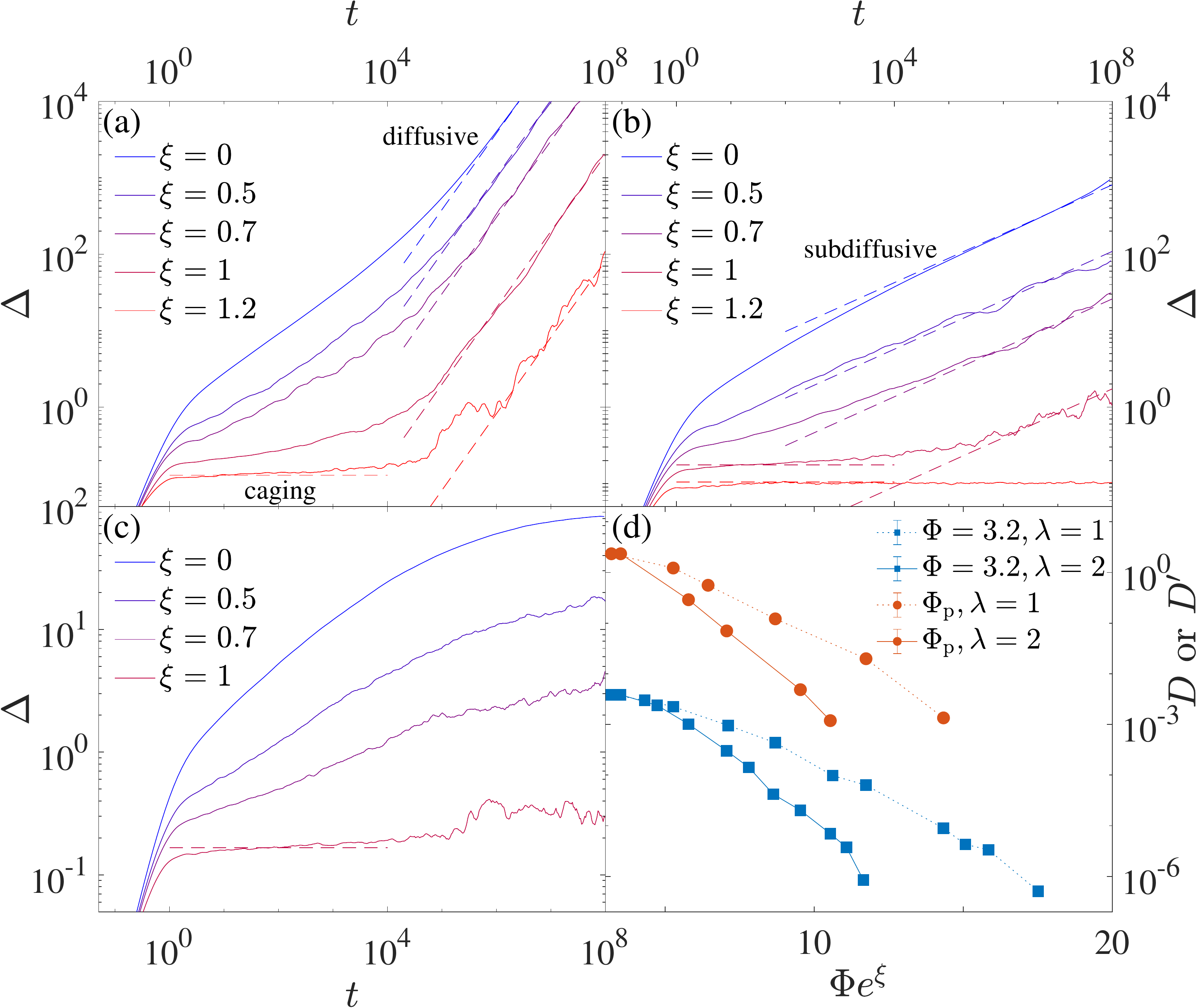}
\caption{Tracer dynamics in the inhomogeneous RLG given by Eq.~\eqref{eq:sqwell} with $\lambda=2$ and various $\xi$ for (a) $\Phi=3.2 < \Phi_\mathrm{p}$, (b) $\Phi=\Phi_\mathrm{p}$, and (c) $\Phi=3.6 > \Phi_\mathrm{p}$. Inclined dashed lines denote the long-time diffusion and subdiffusion regimes in (a) and (b), respectively. Horizontal dashed lines denote the height of long-persisting plateau given by the cage size in homogeneous RLG with $\Phi'=\Phi e^{\xi}$. For $\xi=0$ the original RLG model is recovered. (d) Both the apparent diffusion constant $D$ at $\Phi=3.2$ (blue) and the subdiffusion prefactor $D'$ at $\Phi=\Phi_\mathrm{p}$ (red) shrink with increasing $\xi$ faster than the exponential decay of $e^{-\Phi\exp(\xi)}$.}
\label{fig:inhdyn}
\end{figure}

Motivated by the above results, we consider the impact of the local distribution of obstacles on the interplay between caging and percolation physics. Increasing caging by reaching higher spatial dimensions is out of computational reach, hence we instead consider a three-dimensional systems modified so as to enhance caging. The inhomogeneous RLG has obstacles distributed around the initial tracer position following the spatial probability distribution described by  Eq.~\eqref{eq:sqwell}. The inner shell with an excess density of obstacles around the origin is then expected to make caging more prominent at intermediate times. But because the obstacle density remains unchanged beyond this inner shell, the percolation threshold, which is here defined as the onset $\Phi$ where $\Delta$ diverges in the long-time limit, is not expected to change.

Figure~\ref{fig:inhdyn} illustrates the impact of increasing the local obstacle density on three different regimes of percolation physics.
\begin{enumerate}[label=(\alph*)]
\item For $\Phi=3.2 < \Phi_\mathrm{p}$, enhanced local caging slows the intermediate-time dynamics. The resulting long-time diffusion constant $D = \lim_{t \rightarrow \infty} \Delta/t$ thus decreases as $\xi$ grows.
\item For $\Phi=\Phi_\mathrm{p}$, enhanced local caging also slows the intermediate-time dynamics, but leaves the subdiffusion scaling unchanged. Only the prefactor to that scaling, $D' = \lim_{t \rightarrow \infty} \Delta/t^{2/d'_\mathrm{w}}$, decreases as $\xi$ grows.
\item For $\Phi=3.6 > \Phi_\mathrm{p}$, the long-time cage size shrinks as $\xi$ increases. 
\end{enumerate}
As expected, a dense inner shell affects the intermediate-time dynamics, but leaves the percolation threshold $\Phi_\mathrm{p} = 3.510$ unchanged.

For $\Phi \le \Phi_\mathrm{p}$, the diffusion and subdiffusion prefactors diminish fairly abruptly. The precise decay form is not here determined, but Fig.~\ref{fig:inhdyn}(d) shows that decrease to be faster than exponential with $\Phi e^\xi$. Consequently, cage escapes fall out of the accessible simulation time for $\xi\gtrsim 1$. 
No sharp transition is however observed. $\Delta$ must still diffuse (or subdiffuse) in the infinite-time limit, because the probability that a cavity reaches beyond the inner shell is nonzero. A long-persisting ``intermediate time'' plateau can thus be observed for high enough $\xi$, e.g., $\xi=1.2$ in Fig.~\ref{fig:inhdyn}(a,b).
The height of the long-persisting plateau is given by the long-time limit of the MSD for an homogeneous RLG where $\Phi' = \Phi e^{\xi}$ 
This large $\xi$ regime gives an effective obstacle density well above the percolation threshold, where mean-field theory is known to give robust predictions for the scaling of the cage size~\cite{biroli2020unifying}. Changing $\xi$ therefore provides a continuous way to tune from a mean-field-like to a percolation-like regime, while remaining in physical $d=3$. 

We note that the temperature dependence of these constants on the obstacle distribution is also encoded in $\xi$, and hence this behavior is reminiscent of the super-Arrhenius scaling observed in fragile glasses~\cite{berthier2011theoretical}. It is remarkable that such a modest modification in RLG can apparently capture such characteristic features of glasses. Pursuing this analogy further is, however, left as future work.

\section{Conclusion}  \label{sec:conclusion}

We have obtained the void percolation threshold of the RLG using both an asymptotic scaling analysis and numerics up to $d=9$. 
The numerically determined thresholds suggest a dimensional scaling between $\Phi \sim d$ and $d \log d$, which falls between the conjectured bounds, $\Omega(d) \le \Phi_\mathrm{p} \le \OO(d \log d)$. We hope that these results will inspire more formal mathematical derivations of (possibly tighter) bounds.
We have also examined the percolation criticality beyond $d=3$ using advanced simulation techniques and extensive computations.
We confirm that the geometric critical exponents, including $\tau$, $\gamma$ and $\mu_-$ are identical with those of lattice models, which is consistent with the physical expectation that void percolation belongs to the simple percolation universality class. We also find that the diffusion exponent $\mu$ in $d=4,5$ and the subdiffusion exponent $d'_\mathrm{w}$ in $d=4$ are consistent with the theoretical prediction of Machta~\emph{et al}.~\cite{machta1985diffusion}.

Interestingly, for $d \ge 8$ an additional intermediate-time dynamical slowdown is found to intervene. This phenomenon is absent in lattice percolation, but is a known characteristic of mean-field caging in structural glasses. Mean-field-like caging thus seem to grow more significant and percolation physics to be correspondingly eclipsed as $d$ increases, as reported in Refs.~\onlinecite{biroli2020unifying,biroli2020mean}. This finding motivated our consideration of a modified version of the $d=3$ RLG that tunes the spatial profile of the obstacle distribution. This model, which enhances the intermediate-time dynamical slowdown while keeping the percolation transition unchanged,  further motivates the importance of the mean-field scenario, and offers interesting parallels with structural glass formers. In summary, our work clarifies various aspects of the void percolation and of the interplay between percolation and glass physics in the RLG.

\begin{acknowledgments}
We thank F.~Baccelli, G.~Biroli, E.I.~Corwin, J.~Machta and M.~Teillaud for stimulating discussions and suggestions.
This work was supported by Canada's NSERC (to B.C.) and a grant from the Simons Foundation (\#454937 to P.C.).
The computations were carried out on the Duke Compute Cluster and Open Science Grid~\cite{osg07,osg09}, supported by National Science Foundation award 1148698, and the U.S. Department of Energy's Office of Science.
Data relevant to this work have been archived and can be accessed at Duke digital repository~\cite{lpdata}.
\end{acknowledgments}

\appendix

\section{Void percolation upper bound} \label{appd:phipup}
In this appendix we detail the derivation of the asymptotic upper bound for void percolation following the formalism outlined in Section~\ref{sec:phipup}. Recall that all areas and volumes are here scaled by $\sigma^{d-1}$ and $\sigma^d$ for convenience.

\subsection{Area arithmetic}
We first analyze the expression for $A_\mathrm{s}$ and $A_\mathrm{h}$ for general $d$ (Eq.~\eqref{eq:Pholeexist}). From the construction in Fig.~\ref{fig:geometry}(a), the area of the smallest relevant hole on a shell of radius $r_\mathrm{s}$ can be described as a spherical cap with base radius $r_\mathrm{tracer}$. Its area is~\cite{li2011concise},
\begin{equation} \label{eq:acap}
A_\mathrm{h} = \frac{1}{2} A_d r_\mathrm{s}^{d-1} I_{z_0}\left(\frac{d-1}{2}, \frac{1}{2} \right),
\end{equation}
where $A_d = d V_d$ is the area of $d$-dimensional unit sphere, $I_z(a,b)$ is the regularized incomplete beta function and
\begin{equation}
z_0 = \frac{r_\mathrm{tracer}^2}{r_\mathrm{s}^2} = \frac{ ( 1 - r_\mathrm{obs})^2}{r_\mathrm{s}^2}.
\end{equation}

We also have that the expected total void area on the shell (Fig.~\ref{fig:geometry}(b,c)) is
\begin{equation}
A_\mathrm{s} = A_d r_\mathrm{s}^{d-1} P_\mathrm{s}(r_\mathrm{s}),
\end{equation}
where $P_\mathrm{s}$ is the (conditional) probability that a point at distance $r_\mathrm{s}$ from the origin lies in the vacant space. The probability that no obstacle center is found the shaded area of Fig.~\ref{fig:geometry}(b,c) is thus
\begin{equation} \label{eq:Psdef}
P_\mathrm{s} = \left\{\begin{alignedat}{2}
&\ee^{-\rho (V_2 - V_1)}, & 1 - r_\mathrm{obs} < r_\mathrm{s} \le \sqrt{1 - r_\mathrm{obs}^2} \\ & & \text{(Case 1)}, \\
&\ee^{-\rho (V_d r_\mathrm{obs}^d - V_1 - V_2)},\quad & \sqrt{1 - r_\mathrm{obs}^2} < r_\mathrm{s} <  1 + r_\mathrm{obs}\\ & & \text{(Case 2)}, \\
&\ee^{-\rho V_d r_\mathrm{obs}^d}, & r_\mathrm{s} \ge  1 + r_\mathrm{obs}\\ & & \text{(Case 3)},
\end{alignedat}\right.
\end{equation}
where $V_1$ and $V_2$ are the volumes of the spherical caps on the spheres centered at O and at A with radii OE and AE, respectively [Fig.~\ref{fig:geometry}(b, c)], that share the same base with radius ($\mathrm{PE}$), such that
\begin{equation} \begin{aligned}
r_\mathrm{cap} = \frac{1}{2 r_\mathrm{s}} \sqrt{(1+r_\mathrm{s}+r_\mathrm{obs})(1+r_\mathrm{s}-r_\mathrm{obs})} \cdot \\
\sqrt{(1-r_\mathrm{s}+r_\mathrm{obs})(-1+r_\mathrm{s}+r_\mathrm{obs})}.
\end{aligned} \end{equation}
Given the volume of $d$-dimensional spherical caps for a sphere of radius $r$ and base radius $r_\mathrm{cap}$~\cite{li2011concise}
\begin{equation} \label{eq:vcap}
V_\mathrm{cap} = \frac{1}{2} V_d r^d I_{r_\mathrm{cap}^2/r^2} \left( \frac{d+1}{2}, \frac{1}{2} \right),
\end{equation}
allows us to rewrite Eq.~\eqref{eq:Psdef} as
\begin{equation}
P_\mathrm{s} = \ee^{-\Phi r_\mathrm{obs}^d f_\mathrm{s}(r_\mathrm{s}; r_\mathrm{obs}) } ,
\end{equation}
with the ratio of shaded volume over the whole sphere (AE) volume,
\begin{equation} \label{eq:fsdef}
\begin{aligned}
&f_\mathrm{s}(r_\mathrm{s}; r_\mathrm{obs}) = \\
&\begin{cases}
\frac{1}{2} \left[ I_{z_2} \bigl( \frac{d+1}{2}, \frac12 \bigl) - r_\mathrm{obs}^{-d} I_{z_1} \bigl( \frac{d+1}{2}, \frac12 \bigl) \right], & \text{Case 1}, \\
1 - \frac{1}{2} \left[ I_{z_2} \bigl( \frac{d+1}{2}, \frac12 \bigl) + r_\mathrm{obs}^{-d} I_{z_1} \bigl( \frac{d+1}{2}, \frac12 \bigl) \right], & \text{Case 2}, \\
1, & \text{Case 3}, \\
\end{cases}
\end{aligned}
\end{equation}
where
\begin{align}
z_1 &= \left(\frac{r_\mathrm{cap}}{r}\right)^2 = \frac{1+r_\mathrm{obs}^2}{2} - \frac{(r_\mathrm{obs}^2 - 1)^2}{4 r_\mathrm{s}^2} - \frac{r_\mathrm{s}^2}{4},\label{eq:z1def}\\
z_2 &= \left(\frac{r_\mathrm{cap}}{r_\mathrm{obs}}\right)^2 = \frac{1}{2} + \frac{2 - r_\mathrm{obs}^2}{4 r_\mathrm{s}^2} + \frac{2 - r_\mathrm{s}^2}{4 r_\mathrm{obs}^2} - \frac{1}{4 r_\mathrm{obs}^2 r_\mathrm{s}^2}.\label{eq:z2def}
\end{align}
The area ratio in Eq.~\eqref{eq:Pholeexist} can thus be written as
\begin{equation}
\begin{aligned}
\frac{A_\mathrm{s}}{A_\mathrm{h}} &= \frac{A_d r_\mathrm{s}^{d-1} \ee^{-\Phi r_\mathrm{obs}^d f_\mathrm{s} }}{A_d r_\mathrm{s}^{d-1} I_\mathrm{z_0} \bigl( \frac{d-1}{2}, \frac12 \bigl)/2 }\\
&= 2 \ee^{- \left[\Phi r_\mathrm{obs}^d f_\mathrm{s} + \ln I_\mathrm{z_0} \bigl( \frac{d-1}{2}, \frac12 \bigl) \right] } \\
&\equiv 2 \exp \left[ - F_\mathrm{s}(r_\mathrm{s}; r_\mathrm{obs}; \Phi, d) \right],
\end{aligned}
\end{equation}
where the function $F_\mathrm{s}$ is implicitly defined. 
If $F_\mathrm{s}$ diverges (with any order that grows with $d$) in the limit $d\rightarrow\infty$, then equivalently $A_\mathrm{s}/A_\mathrm{h}$ vanishes. 
If $F_\mathrm{s} = \OO(1)$ then $A_\mathrm{s}/A_\mathrm{h}$ persists and $P_\mathrm{h}$ may not vanish. (Recall that the inequality in Eq.~\eqref{eq:Pholeexist} may not be tight.) These two regimes are separated by a saddle point with $F_\mathrm{s} = \Omega(1)$ under the scaling of $\Phi \sim f_\mathrm{u}(d)$. 

\subsection{Asymptotic scaling analysis}

A standard approach for obtaining a bound on $F_\mathrm{s} = \Omega(1)$ with $d \rightarrow \infty$ is to find $(r_\mathrm{s}, r_\mathrm{obs})$ which maximizes $F_\mathrm{s}$ for a given $d$ and $\Phi$, and identify the dimensional scaling of $\Phi \sim f_\mathrm{u}(d)$, where $F_\mathrm{s} = \Omega(1)$ in the limit $d \rightarrow \infty$. 
Here, we follow a different approach. We first propose a choice of $(r_\mathrm{s}, r_\mathrm{obs})$, and then demonstrate its optimality by showing that it gives the lowest-order growth of $\Phi$. 
(Through this section and next ones, we use the asymptotic notation $g(d)\sim h(d)$ quite loosely to mean $\lim_{d\to\infty} g(d)/h(d)$ is finite and non-zero. This helps us avoid carrying constants.)

In particular, motivated to achieve the optimum at the boundary of the piece-wise function Eq.~\eqref{eq:Psdef}, we propose $r_\mathrm{s} = \sqrt{1-r_\mathrm{obs}^2}$. This choice maximizes the base radius of spherical caps in Fig.~\ref{fig:geometry}, such that $r_\mathrm{cap} = r_\mathrm{obs}$. We then evaluate how the bound on $\Phi$ evolves with $d$ under different $r_\mathrm{obs}$ cases, and find the smallest $f_\mathrm{u}(d)$ among them.

The expression for $F_\mathrm{s}$ involves a couple of regularized incomplete beta functions. We here briefly review these functions, and simplify expressions for $F_\mathrm{s}$ in the asymptotic $d\rightarrow \infty$ limit. This analysis allows us to identify the maximum of $F_\mathrm{s}$ with respect to $(r_\mathrm{s}, r_\mathrm{obs})$ in that limit.
Regularized incomplete beta functions are defined as
\begin{equation}
I_z(a, b) = \frac{B_z(a, b)}{B(a, b)} = \frac{B_z(a, b)\mathrm{\Gamma}(a+b)}{\mathrm{\Gamma}(a)\mathrm{\Gamma}(b)}
\end{equation}
where $B_z(a, b)$ is the incomplete beta function, $B(a, b) = B_1(a, b)$ is the complete beta function, and $\mathrm{\Gamma}(a)$ is the gamma function. For fixed $a, b > 0$, 
$I_z(a,b)$ monotonically grows over the interval $z \in [0, 1]$ and in particular, $I_0(a,b)=0$ and $I_1(a,b)=1$. In the context of Eqs.~\eqref{eq:acap} and~\eqref{eq:vcap}, we seek, for $b=1/2$, $z \in (0, 1)$, asymptotic scaling forms for $a\rightarrow\infty$ of $I_z(a, 1/2)$. From Ref.~\onlinecite[Eq. (2)]{nemes2016uniform}, we have
\begin{equation} \label{eq:uniformbetareg}
I_z(a, 1/2) = \frac{\Gamma(\frac{1}{2}, -a \ln z)}{B(a, \frac{1}{2})\sqrt{a}} \left[ \left(\frac{-\ln z}{1-z}\right)^{\frac{1}{2}} + \OO(a^{-1})\right],
\end{equation}
where $\Gamma(b, c)$ is the incomplete gamma function.

In the derivation below, we provide different asymptotic scaling forms of $I_z(a, 1/2)$ depending on how the function $1/(1-z)$ acts with $a$. 
\begin{enumerate}
\item For $a \gg 1/(1-z) > 0$, expanding the gamma and beta functions around $+\infty$ in Eq.~\eqref{eq:uniformbetareg} gives
\begin{equation} \label{eq:Izapprox1} \begin{aligned}
 &I_z(a, 1/2)  = \frac{z^a}{\sqrt{\pi(1-z)a}} (1 -  \OO [a(1-z)]^{-1} ).
\end{aligned} \end{equation}

\item For $a \sim 1/(1-z) \gg 1$, $I_z$ converges to a nonzero finite value. Denoting the scaled variable $a' = a(1-z)$ and taking $a \rightarrow \infty$ (equivalently, $z \rightarrow 1^-$), 
\begin{equation} \label{eq:Izapprox2}
I_z\left(a, 1/2\right) = \frac{\Gamma(\frac{1}{2}, a')}{\sqrt{\pi}}[1 + \OO(a^{-1}) ].
\end{equation}

\item For $1\ll a \ll 1/(1-z)$, i.e., $a' = a(1-z) \ll 1$, expanding the gamma function around 0 and the beta function around $+\infty$ in Eq.~\eqref{eq:uniformbetareg} gives
\begin{equation} \label{eq:Izapprox3}
\begin{aligned}
1 - I_z(a, 1/2) = 2 \sqrt{ \frac{2 a'}{ \pi } } [1 + \OO(\frac{1}{a})] + \OO(\frac{1}{a}) + \OO(a'^{\frac{3}{2} }).
\end{aligned} \end{equation}

\end{enumerate}

\subsubsection{Evaluation of $f_\mathrm{s}$} \label{sec:fs}

The function $f_\mathrm{s}$ is bounded over the range $f_\mathrm{s} \in (0, 1]$ and grows monotonically with $r_\mathrm{s}$. At $r_\mathrm{s} = \sqrt{1 - r_\mathrm{obs}^2}$, in particular,
\begin{equation}
f_\mathrm{s} = \frac{1}{2}\left[ 1 - r_\mathrm{obs}^{-d} I_{r_\mathrm{obs}^2}\bigl( \frac{d+1}{2}, \frac{1}{2} \bigl) \right].
\end{equation}
Inserting Eq.~\eqref{eq:Izapprox1}~to~\eqref{eq:Izapprox3} gives:
\begin{enumerate}
\item for $d \gg \frac{2}{1 - r_\mathrm{obs}^2} - 1$,
\begin{equation} \label{eq:fscase1}
f_\mathrm{s} \approx \frac12;
\end{equation}

\item for $a' = \frac{d+1}{2}{(1 - r_\mathrm{obs}^2)} = \OO(1)$,
\begin{equation} \label{eq:fscase2}
f_\mathrm{s} \approx \frac{1}{2}\left[ 1 - \left(r_\mathrm{obs}^{-\frac{2 a'}{1-r_\mathrm{obs}^2}+1}\right)  \frac{\Gamma(\frac{1}{2}, a')}{\sqrt{\pi}} \right] = \OO(1);
\end{equation}

\item for $d \ll \frac{2}{1 - r_\mathrm{obs}^2} - 1$,
\begin{equation} \label{eq:fscase3}
\begin{aligned}
f_\mathrm{s} &\approx \frac{1}{2} \left\{ 1 - \Bigl[ 1 + \frac{d-1}{2} (1-r_\mathrm{obs}^2 ) \Bigl]\right. \\
&\quad\left.\Bigl[ 1 - \sqrt{\frac{2(1-r_\mathrm{obs}^2)(d+1)}{\pi}} \Bigl] \right\}\\
&\approx \sqrt\frac{2(1-r_\mathrm{obs}^2)(d+1)}{\pi}.
\end{aligned} 
\end{equation}

\end{enumerate}

\subsubsection{Evaluation of $I_{z_0}$}

The function $I_{z_0} = I_{ (1-r_\mathrm{obs})^2/r_\mathrm{s}^2 }$ monotonically decreases with $r_\mathrm{s}$. At $r_\mathrm{s} = \sqrt{1 - r_\mathrm{obs}^2}$, specifically, $z_0 = (1-r_\mathrm{obs})/(1+r_\mathrm{obs})$. Unless $1/(1-z_0) \gtrsim (d-1)/2$, i.e., $1/r_\mathrm{obs} \ll d$, in the large $d$ limit we can apply Eq.~\eqref{eq:Izapprox1} to obtain
\begin{equation} \label{eq:Iz0asymp}
\begin{aligned}
\ln &I_{z_0} \left(\frac{d-1}{2}, \frac12\right) \approx \ln \left[ \left(\frac{1-r_\mathrm{obs}}{1+r_\mathrm{obs}}\right)^{\frac{d-1}{2}} \sqrt{\frac{1 + r_\mathrm{obs}}{\pi (d-1) r_\mathrm{obs} }} \right] \\
&= \frac{d-1}{2} \ln(1-r_\mathrm{obs}) -\frac{d-2}{2} \ln (1+r_\mathrm{obs}) \\
&\quad -\frac{1}{2} \ln(d-1) - \frac{1}{2} \ln(\pi r_\mathrm{obs}),
\end{aligned} 
\end{equation}
where the four terms on the right-hand size grow as $\Omega (d), \OO (d), \OO (\log d)$ and $\OO (1)$, respectively. Dropping the sub-leading terms as $d\rightarrow\infty$ then gives
\begin{equation} \label{eq:Iz0approx}
\ln I_{z_0} \left(\frac{d-1}{2}, \frac12\right) \approx \frac{d-1}{2} \ln(1-r_\mathrm{obs}).
\end{equation}
Before going further we shall check the resulting percolation upper bound for the case $1/r_\mathrm{obs} \gtrsim d$ for a moment, for which $\ln I_{z_0}$ can no longer be approximated by Eq.~\eqref{eq:Iz0approx}. 
Because $f_\mathrm{s} \le 1$ and $I_{z_0} < 1$, we always have $F_\mathrm{s} < \Phi r_\mathrm{obs}^d$. To make $F_\mathrm{s} = \Omega(1)$ we need $\Phi = \Omega( (1 / r_\mathrm{obs}) ^{d}) \ge \Omega( d^d) $, which grows super-exponentially with $d$. 
Ruling out this case, in the following we only consider $1/r_\mathrm{obs} < d$, for which Eq.~\eqref{eq:Iz0approx} can always be used to approximate $I_{z_0}$. This consideration greatly simplifies the discussion below. 

\subsection{Bound determination}

With the asymptotic expressions of $f_\mathrm{s}$, we can now obtain the condition for $F_\mathrm{s} = \Omega(1)$ with $d \rightarrow \infty$ under different choices of $r_\mathrm{obs}$.

\begin{enumerate}
\item For  $d \gg \frac{2}{1 - r_\mathrm{obs}^2} - 1$, we have
\begin{equation} \begin{aligned}
F_s &\approx \Phi r_\mathrm{obs}^d \cdot \frac{1}{2} + \frac{d-1}{2} \ln ( 1-r_\mathrm{obs})  \\
&\le \Phi r_\mathrm{obs}^d \cdot \frac{1}{2} - \frac{C_1}{2} d \\
&\approx \frac{1}{2} \left( \Phi r_\mathrm{obs}^d - C_1 d \right).
\end{aligned} \end{equation}
where $C_1$ is a positive constant of order unity. To make $F_\mathrm{s} = \Omega(1)$, $\Phi$ needs to grows at least as $\Phi = \Omega ( (1/ r_\mathrm{obs} )^d ) $, which is a super-exponential bound.

\item For $a' = \frac{d+1}{2}{(1 - r_\mathrm{obs}^2)} = \OO(1)$, we have that $\lim_{d \rightarrow \infty} r_\mathrm{obs}^d = \ee^{-a'}$ is a constant, and thus
\begin{equation} \begin{aligned}
F_\mathrm{s} &\approx \Phi \cdot \ee^{-a'} f_s + \frac{d-1}{2 } \ln ( 1-r_\mathrm{obs}) \\
&\approx \Phi \cdot C_2 - \frac{d-1}{2 } \ln \left( \frac{d+1}{a'} \right) \\
&\approx \Phi \cdot C_2 - d \ln d /2,
\end{aligned} \end{equation}
with $C_2 =  \ee^{-a'} f_s$ a positive constant of order unity. To make $F_\mathrm{s} = \Omega(1)$, we need at least $\Phi = \Omega(d \log d)$. 

\item In the case  $a'=o(1)$, we have that $d \ll \frac{2}{1 - r_\mathrm{obs}^2} - 1$ then $r_\mathrm{obs}^d \approx 1$, and $f_\mathrm{s}$ also follows Eq.~\eqref{eq:fscase3} and vanishes in the limit of $d \rightarrow \infty$. Therefore,
\begin{equation} \begin{aligned}
F_\mathrm{s} &\approx \Phi f_\mathrm{s} + \frac{d-1}{2} \ln ( 1 - r_\mathrm{obs} ) \\
&\approx \Phi f_\mathrm{s} - \frac{d-1}{2} \left[ \ln ( d+1) + \ln (1+r_\mathrm{obs})\right. \\
&\quad \left. -2 \ln f_\mathrm{s} - \ln (\pi/2) \right]. \\
\end{aligned} \end{equation}
and we need that $F_\mathrm{s} = \Omega(1)$.  
If $f_\mathrm{s}$ vanishes under any polynomial order $f_\mathrm{s} \sim d^{-n}$ with $n>0$, then we need $\Phi \sim d^{1+n} \log d$. In the limit of $n \rightarrow 0$ the previous scaling is recovered. If $f_\mathrm{s}$ vanishes faster than polynomial, then $\Phi \sim d \log f / f_\mathrm{s}(d)$. For example, if $f_\mathrm{s} \sim \ee^{-d}$, $\Phi \sim d^2 \ee^d $. Because $-\log f_\mathrm{s} > \log d$ and $1/f_\mathrm{s} = \Omega(1)$, however, $\Phi \sim d \log f / f_\mathrm{s}(d)>d \log d$.

\end{enumerate} 

In summary, we have shown that for $r_\mathrm{s} = \sqrt{1 - r_\mathrm{obs}^2}$, the tightest bound for $F_\mathrm{s} = \Omega(1)$ in the asymptotic $d\rightarrow\infty$ is $\Phi \sim d \log d$, and that this bound is achieved by choosing $\frac{d+1}{2}{(1 - r_\mathrm{obs}^2)} = \OO(1)$, or, in other words, that $r_\mathrm{obs} \sim \sqrt{\frac{d-1}{d+1}}$. Under this choice of $(r_\mathrm{s}, r_\mathrm{obs})$, $P_\mathrm{h}$ vanishes in $\Phi = \Omega(d \log d)$. Because this condition is sufficient for void percolation not to occur, we have therefore obtained an upper bound for the void percolation threshold,
\begin{equation}
\Phi_\mathrm{p}(d) = \OO(d \log d).
\end{equation}

\subsection{Bound validation} \label{sec:upconfirm}

As a complement, we here confirm that the bound obtained by choosing $r_\mathrm{s} = r^*_\mathrm{s} = \sqrt{1 - r_\mathrm{obs}^2}$ is optimal.
Again, denoting $ a' = \frac{d+1}{2}(1-r_\mathrm{obs}^2) = \OO(1)$ and $r_\mathrm{s} = r_\mathrm{s}^*(1+\epsilon)$, we have
\begin{align}
F_\mathrm{s} &\approx \Phi \ee^{-a'} f_\mathrm{s} + \ln I_{z_0}.
\end{align}
Because $F_\mathrm{s}$ depends on the scalings of both $f_\mathrm{s}$ and $\ln I_\mathrm{z_0}$, we propose the following strategy: (i) when $\ln I_\mathrm{z_0} = \OO(1)$, we show that $f_\mathrm{s} = \OO(1/(d \log d))$; (ii) when $f_\mathrm{s} = \OO(1)$, we show that $\ln I_{z_0} = - \Omega(d \log d)$. Then, $F_\mathrm{s} = \Omega(1)$ only when $\Phi = \Omega(d \log d)$.

\begin{enumerate}

\item When $r_\mathrm{tracer} < r_\mathrm{s} < r^*_\mathrm{s}$ (case 1 in Eq.~\eqref{eq:fsdef}), i.e., $-1+\sqrt{z_0^*} < \epsilon < 0$. 

For notational convenience, we write $1+\epsilon = b'  \sqrt{z_0^*}$, where $1 < b' < 1/\sqrt{z_0^*} \approx \sqrt{(d+1)/a'}$.
Inserting $r_\mathrm{s} = r^*_\mathrm{s}  b' \sqrt{z_0^*}$ into Eq.~\eqref{eq:z2def} gives
\begin{equation} \begin{aligned}
z_2 &= \frac{1}{4}(1-\frac{1}{b'^2}) \left[ \left(1+\frac{1}{r_\mathrm{obs}}\right)^2 - b'^2 \left(1-\frac{1}{r_\mathrm{obs}}\right)^2 \right] \\
&= 1 - \frac{1}{b'^2} + \OO(\frac{1}{d})
\end{aligned} \end{equation}
Assuming $b' \le \sqrt{d}/\ln d$, which satisfies $b'^2 \ll \frac{d+1}{2}$, so we can apply Eq.~\eqref{eq:Izapprox1} and then Eq.~\eqref{eq:fsdef} (Case 1),
\begin{equation}
f_\mathrm{s} < \frac{1}{2} I_{z_2} (\frac{d+1}{2}, \frac{1}{2}) \sim \frac{b'}{\sqrt{d+1}} \exp(-\frac{d+1}{2 b'^2}).
\end{equation}
Then,
\begin{equation}
F_\mathrm{s} < \Phi e^{-a'} \cdot f_\mathrm{s} \le \frac{\Phi e^{-a'}}{\ln d \cdot d^{\ln d/2} } <  \frac{\Phi e^{-a'}}{d \ln d }
\end{equation}
A necessary condition for $F_\mathrm{s} = \Omega(1)$ is then $\Phi = \Omega(d \log d)$. 

When $b' > \sqrt{d}/\ln d$, we simply use the fact that $f_\mathrm{s} \sim \OO(1)$ and investigate the scaling on $\ln I_{z_0}$. 
Since $z_0 = z^*_0/(1+\epsilon)^2 = 1/b'^2 \ll 1$, we can again apply Eq.~\eqref{eq:Izapprox1} and obtain
\begin{equation} \begin{aligned}
\ln I_{z_0} &= -\frac{1}{2} \ln \left[ \pi \left(1 - \frac{1}{b'^2}\right)\frac{d-1}{2} \right] - \frac{d-1}{2} \ln (b'^2) \\
&< - \frac{d-1}{2} \ln \left( \frac{d}{(\ln d)^2}\right) \\
&= -\frac{d-1}{2} \ln d + (d-1) \ln \ln d,
\end{aligned} \end{equation}
where the second term is sub-dominant, and hence $\ln I_\mathrm{z_0} = -\Omega(d \log d)$. Therefore, for $F_\mathrm{s} = \Omega(1)$, $\Phi$ should grow at least as $\Phi = \Omega(d \log d)$.

\item $r_\mathrm{s} > r^*_\mathrm{s}$ (cases 2, 3 in Eq.~\eqref{eq:fsdef}), i.e. $\epsilon > 0$. Because $0 < f_\mathrm{s}^* < f_\mathrm{s} < 1$, and $I_\mathrm{z_0} < I_\mathrm{z_0^*}$, we have
\begin{equation} \begin{aligned}
F_\mathrm{s} &<  \Phi \cdot \ee^{-a'}+ \ln I_{z_0^*} \\
&\approx  \Phi \cdot \ee^{-a'} - d \ln d / 2. \\
\end{aligned} \end{equation}  
For $F_\mathrm{s} = \Omega(1)$, $\Phi$ should grow at least as $\Phi = \Omega(d \log d)$.

\end{enumerate}

Combining these two cases, we conclude that the bound $\Phi = \Omega(d \log d)$, obtained by the choice $r_\mathrm{s} = r^*_\mathrm{s}$, is optimal. 

\bibliography{abbrev}

%merlin.mbs apsrev4-1.bst 2010-07-25 4.21a (PWD, AO, DPC) hacked
%Control: key (0)
%Control: author (8) initials jnrlst
%Control: editor formatted (1) identically to author
%Control: production of article title (-1) disabled
%Control: page (0) single
%Control: year (1) truncated
%Control: production of eprint (0) enabled
\begin{thebibliography}{64}%
\makeatletter
\providecommand \@ifxundefined [1]{%
 \@ifx{#1\undefined}
}%
\providecommand \@ifnum [1]{%
 \ifnum #1\expandafter \@firstoftwo
 \else \expandafter \@secondoftwo
 \fi
}%
\providecommand \@ifx [1]{%
 \ifx #1\expandafter \@firstoftwo
 \else \expandafter \@secondoftwo
 \fi
}%
\providecommand \natexlab [1]{#1}%
\providecommand \enquote  [1]{``#1''}%
\providecommand \bibnamefont  [1]{#1}%
\providecommand \bibfnamefont [1]{#1}%
\providecommand \citenamefont [1]{#1}%
\providecommand \href@noop [0]{\@secondoftwo}%
\providecommand \href [0]{\begingroup \@sanitize@url \@href}%
\providecommand \@href[1]{\@@startlink{#1}\@@href}%
\providecommand \@@href[1]{\endgroup#1\@@endlink}%
\providecommand \@sanitize@url [0]{\catcode `\\12\catcode `\$12\catcode
  `\&12\catcode `\#12\catcode `\^12\catcode `\_12\catcode `\%12\relax}%
\providecommand \@@startlink[1]{}%
\providecommand \@@endlink[0]{}%
\providecommand \url  [0]{\begingroup\@sanitize@url \@url }%
\providecommand \@url [1]{\endgroup\@href {#1}{\urlprefix }}%
\providecommand \urlprefix  [0]{URL }%
\providecommand \Eprint [0]{\href }%
\providecommand \doibase [0]{http://dx.doi.org/}%
\providecommand \selectlanguage [0]{\@gobble}%
\providecommand \bibinfo  [0]{\@secondoftwo}%
\providecommand \bibfield  [0]{\@secondoftwo}%
\providecommand \translation [1]{[#1]}%
\providecommand \BibitemOpen [0]{}%
\providecommand \bibitemStop [0]{}%
\providecommand \bibitemNoStop [0]{.\EOS\space}%
\providecommand \EOS [0]{\spacefactor3000\relax}%
\providecommand \BibitemShut  [1]{\csname bibitem#1\endcsname}%
\let\auto@bib@innerbib\@empty
%</preamble>
\bibitem [{\citenamefont {Stauffer}\ and\ \citenamefont
  {Aharony}(1994)}]{stauffer1994percolation}%
  \BibitemOpen
  \bibfield  {author} {\bibinfo {author} {\bibfnamefont {D.}~\bibnamefont
  {Stauffer}}\ and\ \bibinfo {author} {\bibfnamefont {A.}~\bibnamefont
  {Aharony}},\ }\href {\doibase 10.1201/9781315274386} {\emph {\bibinfo {title}
  {Introduction To Percolation Theory}}}\ (\bibinfo  {publisher} {Taylor \&
  Francis},\ \bibinfo {year} {1994})\BibitemShut {NoStop}%
\bibitem [{\citenamefont {Ben-Avraham}\ and\ \citenamefont
  {Havlin}(2000)}]{ben2000diffusion}%
  \BibitemOpen
  \bibfield  {author} {\bibinfo {author} {\bibfnamefont {D.}~\bibnamefont
  {Ben-Avraham}}\ and\ \bibinfo {author} {\bibfnamefont {S.}~\bibnamefont
  {Havlin}},\ }\href {\doibase 10.1017/CBO9780511605826} {\emph {\bibinfo
  {title} {Diffusion and reactions in fractals and disordered systems}}}\
  (\bibinfo  {publisher} {Cambridge University Press},\ \bibinfo {year}
  {2000})\BibitemShut {NoStop}%
\bibitem [{\citenamefont {Berkowitz}\ \emph {et~al.}(2006)\citenamefont
  {Berkowitz}, \citenamefont {Cortis}, \citenamefont {Dentz},\ and\
  \citenamefont {Scher}}]{berkowitz2006modeling}%
  \BibitemOpen
  \bibfield  {author} {\bibinfo {author} {\bibfnamefont {B.}~\bibnamefont
  {Berkowitz}}, \bibinfo {author} {\bibfnamefont {A.}~\bibnamefont {Cortis}},
  \bibinfo {author} {\bibfnamefont {M.}~\bibnamefont {Dentz}}, \ and\ \bibinfo
  {author} {\bibfnamefont {H.}~\bibnamefont {Scher}},\ }\href {\doibase
  10.1029/2005RG000178} {\bibfield  {journal} {\bibinfo  {journal} {Rev.
  Geophys}\ }\textbf {\bibinfo {volume} {44}},\ \bibinfo {pages} {RG2003}
  (\bibinfo {year} {2006})}\BibitemShut {NoStop}%
\bibitem [{\citenamefont {H{\"o}fling}\ and\ \citenamefont
  {Franosch}(2013)}]{hofling2013anomalous}%
  \BibitemOpen
  \bibfield  {author} {\bibinfo {author} {\bibfnamefont {F.}~\bibnamefont
  {H{\"o}fling}}\ and\ \bibinfo {author} {\bibfnamefont {T.}~\bibnamefont
  {Franosch}},\ }\href {\doibase 10.1088/0034-4885/76/4/046602} {\bibfield
  {journal} {\bibinfo  {journal} {Rep. Prog. Phys.}\ }\textbf {\bibinfo
  {volume} {76}},\ \bibinfo {pages} {046602} (\bibinfo {year}
  {2013})}\BibitemShut {NoStop}%
\bibitem [{\citenamefont {Dorogovtsev}\ \emph {et~al.}(2008)\citenamefont
  {Dorogovtsev}, \citenamefont {Goltsev},\ and\ \citenamefont
  {Mendes}}]{dorogovtsev2008critical}%
  \BibitemOpen
  \bibfield  {author} {\bibinfo {author} {\bibfnamefont {S.~N.}\ \bibnamefont
  {Dorogovtsev}}, \bibinfo {author} {\bibfnamefont {A.~V.}\ \bibnamefont
  {Goltsev}}, \ and\ \bibinfo {author} {\bibfnamefont {J.~F.~F.}\ \bibnamefont
  {Mendes}},\ }\href {\doibase 10.1103/RevModPhys.80.1275} {\bibfield
  {journal} {\bibinfo  {journal} {Rev. Mod. Phys.}\ }\textbf {\bibinfo {volume}
  {80}},\ \bibinfo {pages} {1275} (\bibinfo {year} {2008})}\BibitemShut
  {NoStop}%
\bibitem [{\citenamefont {Straley}(1982)}]{straley1982non}%
  \BibitemOpen
  \bibfield  {author} {\bibinfo {author} {\bibfnamefont {J.~P.}\ \bibnamefont
  {Straley}},\ }\href {\doibase 10.1088/0022-3719/15/11/014} {\bibfield
  {journal} {\bibinfo  {journal} {J. Phys. C}\ }\textbf {\bibinfo {volume}
  {15}},\ \bibinfo {pages} {2343} (\bibinfo {year} {1982})}\BibitemShut
  {NoStop}%
\bibitem [{\citenamefont {Halperin}\ \emph {et~al.}(1985)\citenamefont
  {Halperin}, \citenamefont {Feng},\ and\ \citenamefont
  {Sen}}]{halperin1985differences}%
  \BibitemOpen
  \bibfield  {author} {\bibinfo {author} {\bibfnamefont {B.~I.}\ \bibnamefont
  {Halperin}}, \bibinfo {author} {\bibfnamefont {S.}~\bibnamefont {Feng}}, \
  and\ \bibinfo {author} {\bibfnamefont {P.~N.}\ \bibnamefont {Sen}},\ }\href
  {\doibase 10.1103/PhysRevLett.54.2391} {\bibfield  {journal} {\bibinfo
  {journal} {Phys. Rev. Lett.}\ }\textbf {\bibinfo {volume} {54}},\ \bibinfo
  {pages} {2391} (\bibinfo {year} {1985})}\BibitemShut {NoStop}%
\bibitem [{\citenamefont {Xu}\ \emph {et~al.}(2014)\citenamefont {Xu},
  \citenamefont {Wang}, \citenamefont {Lv},\ and\ \citenamefont
  {Deng}}]{xu2014simultaneous}%
  \BibitemOpen
  \bibfield  {author} {\bibinfo {author} {\bibfnamefont {X.}~\bibnamefont
  {Xu}}, \bibinfo {author} {\bibfnamefont {J.}~\bibnamefont {Wang}}, \bibinfo
  {author} {\bibfnamefont {J.-P.}\ \bibnamefont {Lv}}, \ and\ \bibinfo {author}
  {\bibfnamefont {Y.}~\bibnamefont {Deng}},\ }\href {\doibase
  10.1007/s11467-013-0403-z} {\bibfield  {journal} {\bibinfo  {journal} {Front.
  Phys.}\ }\textbf {\bibinfo {volume} {9}},\ \bibinfo {pages} {113} (\bibinfo
  {year} {2014})}\BibitemShut {NoStop}%
\bibitem [{\citenamefont {Mertens}\ and\ \citenamefont
  {Moore}(2018)}]{mertens2018percolation}%
  \BibitemOpen
  \bibfield  {author} {\bibinfo {author} {\bibfnamefont {S.}~\bibnamefont
  {Mertens}}\ and\ \bibinfo {author} {\bibfnamefont {C.}~\bibnamefont
  {Moore}},\ }\href {\doibase 10.1103/PhysRevE.98.022120} {\bibfield  {journal}
  {\bibinfo  {journal} {Phys. Rev. E}\ }\textbf {\bibinfo {volume} {98}},\
  \bibinfo {pages} {022120} (\bibinfo {year} {2018})}\BibitemShut {NoStop}%
\bibitem [{\citenamefont {Zhang}\ \emph {et~al.}(2020)\citenamefont {Zhang},
  \citenamefont {Hou}, \citenamefont {Fang}, \citenamefont {Hu},\ and\
  \citenamefont {Deng}}]{zhang2020critical}%
  \BibitemOpen
  \bibfield  {author} {\bibinfo {author} {\bibfnamefont {Z.}~\bibnamefont
  {Zhang}}, \bibinfo {author} {\bibfnamefont {P.}~\bibnamefont {Hou}}, \bibinfo
  {author} {\bibfnamefont {S.}~\bibnamefont {Fang}}, \bibinfo {author}
  {\bibfnamefont {H.}~\bibnamefont {Hu}}, \ and\ \bibinfo {author}
  {\bibfnamefont {Y.}~\bibnamefont {Deng}},\ }\href@noop {} {\bibfield
  {journal} {\bibinfo  {journal} {arXiv preprint}\ } (\bibinfo {year}
  {2020})},\ \Eprint {http://arxiv.org/abs/arXiv:2004.11289} {arXiv:2004.11289}
  \BibitemShut {NoStop}%
\bibitem [{\citenamefont {Machta}\ and\ \citenamefont
  {Moore}(1985)}]{machta1985diffusion}%
  \BibitemOpen
  \bibfield  {author} {\bibinfo {author} {\bibfnamefont {J.}~\bibnamefont
  {Machta}}\ and\ \bibinfo {author} {\bibfnamefont {S.~M.}\ \bibnamefont
  {Moore}},\ }\href {\doibase 10.1103/PhysRevA.32.3164} {\bibfield  {journal}
  {\bibinfo  {journal} {Phys. Rev. A}\ }\textbf {\bibinfo {volume} {32}},\
  \bibinfo {pages} {3164} (\bibinfo {year} {1985})}\BibitemShut {NoStop}%
\bibitem [{\citenamefont {Havlin}\ and\ \citenamefont
  {Ben-Avraham}(1987)}]{havlin1987diffusion}%
  \BibitemOpen
  \bibfield  {author} {\bibinfo {author} {\bibfnamefont {S.}~\bibnamefont
  {Havlin}}\ and\ \bibinfo {author} {\bibfnamefont {D.}~\bibnamefont
  {Ben-Avraham}},\ }\href {\doibase 10.1080/00018738700101072} {\bibfield
  {journal} {\bibinfo  {journal} {Adv. Phys}\ }\textbf {\bibinfo {volume}
  {36}},\ \bibinfo {pages} {695} (\bibinfo {year} {1987})}\BibitemShut
  {NoStop}%
\bibitem [{\citenamefont {H{\"o}fling}\ \emph {et~al.}(2006)\citenamefont
  {H{\"o}fling}, \citenamefont {Franosch},\ and\ \citenamefont
  {Frey}}]{hofling2006localization}%
  \BibitemOpen
  \bibfield  {author} {\bibinfo {author} {\bibfnamefont {F.}~\bibnamefont
  {H{\"o}fling}}, \bibinfo {author} {\bibfnamefont {T.}~\bibnamefont
  {Franosch}}, \ and\ \bibinfo {author} {\bibfnamefont {E.}~\bibnamefont
  {Frey}},\ }\href {\doibase 10.1103/PhysRevLett.96.165901} {\bibfield
  {journal} {\bibinfo  {journal} {Phys. Rev. Lett.}\ }\textbf {\bibinfo
  {volume} {96}},\ \bibinfo {pages} {165901} (\bibinfo {year}
  {2006})}\BibitemShut {NoStop}%
\bibitem [{\citenamefont {H{\"o}fling}\ \emph {et~al.}(2008)\citenamefont
  {H{\"o}fling}, \citenamefont {Munk}, \citenamefont {Frey},\ and\
  \citenamefont {Franosch}}]{hofling2008critical}%
  \BibitemOpen
  \bibfield  {author} {\bibinfo {author} {\bibfnamefont {F.}~\bibnamefont
  {H{\"o}fling}}, \bibinfo {author} {\bibfnamefont {T.}~\bibnamefont {Munk}},
  \bibinfo {author} {\bibfnamefont {E.}~\bibnamefont {Frey}}, \ and\ \bibinfo
  {author} {\bibfnamefont {T.}~\bibnamefont {Franosch}},\ }\href {\doibase
  10.1063/1.2901170} {\bibfield  {journal} {\bibinfo  {journal} {J. Chem.
  Phys.}\ }\textbf {\bibinfo {volume} {128}},\ \bibinfo {pages} {164517}
  (\bibinfo {year} {2008})}\BibitemShut {NoStop}%
\bibitem [{\citenamefont {Biroli}\ \emph
  {et~al.}(2021{\natexlab{a}})\citenamefont {Biroli}, \citenamefont
  {Charbonneau}, \citenamefont {Corwin}, \citenamefont {Hu}, \citenamefont
  {Ikeda}, \citenamefont {Szamel},\ and\ \citenamefont
  {Zamponi}}]{biroli2020unifying}%
  \BibitemOpen
  \bibfield  {author} {\bibinfo {author} {\bibfnamefont {G.}~\bibnamefont
  {Biroli}}, \bibinfo {author} {\bibfnamefont {P.}~\bibnamefont {Charbonneau}},
  \bibinfo {author} {\bibfnamefont {E.~I.}\ \bibnamefont {Corwin}}, \bibinfo
  {author} {\bibfnamefont {Y.}~\bibnamefont {Hu}}, \bibinfo {author}
  {\bibfnamefont {H.}~\bibnamefont {Ikeda}}, \bibinfo {author} {\bibfnamefont
  {G.}~\bibnamefont {Szamel}}, \ and\ \bibinfo {author} {\bibfnamefont
  {F.}~\bibnamefont {Zamponi}},\ }\href {\doibase 10.1103/PhysRevE.103.L030104}
  {\bibfield  {journal} {\bibinfo  {journal} {Phys. Rev. E}\ }\textbf {\bibinfo
  {volume} {103}},\ \bibinfo {pages} {L030104} (\bibinfo {year}
  {2021}{\natexlab{a}})}\BibitemShut {NoStop}%
\bibitem [{\citenamefont {Meester}\ and\ \citenamefont
  {Roy}(1996)}]{Meester1996}%
  \BibitemOpen
  \bibfield  {author} {\bibinfo {author} {\bibfnamefont {R.}~\bibnamefont
  {Meester}}\ and\ \bibinfo {author} {\bibfnamefont {R.}~\bibnamefont {Roy}},\
  }\href {\doibase 10.1017/CBO9780511895357} {\emph {\bibinfo {title}
  {Continuum Percolation}}},\ \bibinfo {series} {Cambridge Tracts in
  Mathematics}, Vol.\ \bibinfo {volume} {119}\ (\bibinfo  {publisher}
  {Cambridge University Press},\ \bibinfo {address} {Cambridge},\ \bibinfo
  {year} {1996})\BibitemShut {NoStop}%
\bibitem [{\citenamefont {Torquato}(2012)}]{torquato2012effect1}%
  \BibitemOpen
  \bibfield  {author} {\bibinfo {author} {\bibfnamefont {S.}~\bibnamefont
  {Torquato}},\ }\href {\doibase 10.1063/1.3679861} {\bibfield  {journal}
  {\bibinfo  {journal} {J. Chem. Phys.}\ }\textbf {\bibinfo {volume} {136}},\
  \bibinfo {pages} {054106} (\bibinfo {year} {2012})}\BibitemShut {NoStop}%
\bibitem [{\citenamefont {Anantharam}\ \emph {et~al.}(2016)\citenamefont
  {Anantharam}, \citenamefont {Baccelli} \emph
  {et~al.}}]{anantharam2016boolean}%
  \BibitemOpen
  \bibfield  {author} {\bibinfo {author} {\bibfnamefont {V.}~\bibnamefont
  {Anantharam}}, \bibinfo {author} {\bibfnamefont {F.}~\bibnamefont
  {Baccelli}},  \emph {et~al.},\ }\href {\doibase 10.1017/jpr.2016.60}
  {\bibfield  {journal} {\bibinfo  {journal} {J. Appl. Probab.}\ }\textbf
  {\bibinfo {volume} {53}},\ \bibinfo {pages} {1001} (\bibinfo {year}
  {2016})}\BibitemShut {NoStop}%
\bibitem [{\citenamefont {Torquato}\ and\ \citenamefont
  {Jiao}(2012)}]{torquato2012effect2}%
  \BibitemOpen
  \bibfield  {author} {\bibinfo {author} {\bibfnamefont {S.}~\bibnamefont
  {Torquato}}\ and\ \bibinfo {author} {\bibfnamefont {Y.}~\bibnamefont
  {Jiao}},\ }\href {\doibase 10.1063/1.4742750} {\bibfield  {journal} {\bibinfo
   {journal} {J. Chem. Phys.}\ }\textbf {\bibinfo {volume} {137}},\ \bibinfo
  {pages} {074106} (\bibinfo {year} {2012})}\BibitemShut {NoStop}%
\bibitem [{\citenamefont {Jin}\ and\ \citenamefont
  {Charbonneau}(2015)}]{jin2015dimensional}%
  \BibitemOpen
  \bibfield  {author} {\bibinfo {author} {\bibfnamefont {Y.}~\bibnamefont
  {Jin}}\ and\ \bibinfo {author} {\bibfnamefont {P.}~\bibnamefont
  {Charbonneau}},\ }\href {\doibase 10.1103/PhysRevE.91.042313} {\bibfield
  {journal} {\bibinfo  {journal} {Phys. Rev. E}\ }\textbf {\bibinfo {volume}
  {91}},\ \bibinfo {pages} {042313} (\bibinfo {year} {2015})}\BibitemShut
  {NoStop}%
\bibitem [{\citenamefont {G{\"o}tze}\ \emph {et~al.}(1981)\citenamefont
  {G{\"o}tze}, \citenamefont {Leutheusser},\ and\ \citenamefont
  {Yip}}]{gotze1981dynamical}%
  \BibitemOpen
  \bibfield  {author} {\bibinfo {author} {\bibfnamefont {W.}~\bibnamefont
  {G{\"o}tze}}, \bibinfo {author} {\bibfnamefont {E.}~\bibnamefont
  {Leutheusser}}, \ and\ \bibinfo {author} {\bibfnamefont {S.}~\bibnamefont
  {Yip}},\ }\href {\doibase 10.1103/PhysRevA.23.2634} {\bibfield  {journal}
  {\bibinfo  {journal} {Phys. Rev. A}\ }\textbf {\bibinfo {volume} {23}},\
  \bibinfo {pages} {2634} (\bibinfo {year} {1981})}\BibitemShut {NoStop}%
\bibitem [{\citenamefont {Leutheusser}(1984)}]{leutheusser1984dynamical}%
  \BibitemOpen
  \bibfield  {author} {\bibinfo {author} {\bibfnamefont {E.}~\bibnamefont
  {Leutheusser}},\ }\href {\doibase 10.1103/PhysRevA.29.2765} {\bibfield
  {journal} {\bibinfo  {journal} {Phys. Rev. A}\ }\textbf {\bibinfo {volume}
  {29}},\ \bibinfo {pages} {2765} (\bibinfo {year} {1984})}\BibitemShut
  {NoStop}%
\bibitem [{\citenamefont {Szamel}(2004)}]{szamel2004gaussian}%
  \BibitemOpen
  \bibfield  {author} {\bibinfo {author} {\bibfnamefont {G.}~\bibnamefont
  {Szamel}},\ }\href {\doibase 10.1209/epl/i2003-10115-2} {\bibfield  {journal}
  {\bibinfo  {journal} {Europhys. Lett.}\ }\textbf {\bibinfo {volume} {65}},\
  \bibinfo {pages} {498} (\bibinfo {year} {2004})}\BibitemShut {NoStop}%
\bibitem [{\citenamefont {Krakoviack}(2007)}]{krakoviack2007mode}%
  \BibitemOpen
  \bibfield  {author} {\bibinfo {author} {\bibfnamefont {V.}~\bibnamefont
  {Krakoviack}},\ }\href {\doibase 10.1103/PhysRevE.75.031503} {\bibfield
  {journal} {\bibinfo  {journal} {Phys. Rev. E}\ }\textbf {\bibinfo {volume}
  {75}},\ \bibinfo {pages} {031503} (\bibinfo {year} {2007})}\BibitemShut
  {NoStop}%
\bibitem [{\citenamefont {Kim}\ \emph {et~al.}(2009)\citenamefont {Kim},
  \citenamefont {Miyazaki},\ and\ \citenamefont {Saito}}]{kim2009slow}%
  \BibitemOpen
  \bibfield  {author} {\bibinfo {author} {\bibfnamefont {K.}~\bibnamefont
  {Kim}}, \bibinfo {author} {\bibfnamefont {K.}~\bibnamefont {Miyazaki}}, \
  and\ \bibinfo {author} {\bibfnamefont {S.}~\bibnamefont {Saito}},\ }\href
  {\doibase 10.1209/0295-5075/88/36002} {\bibfield  {journal} {\bibinfo
  {journal} {Europhys. Lett.}\ }\textbf {\bibinfo {volume} {88}},\ \bibinfo
  {pages} {36002} (\bibinfo {year} {2009})}\BibitemShut {NoStop}%
\bibitem [{\citenamefont {Kurzidim}\ \emph {et~al.}(2009)\citenamefont
  {Kurzidim}, \citenamefont {Coslovich},\ and\ \citenamefont
  {Kahl}}]{kurzidim2009single}%
  \BibitemOpen
  \bibfield  {author} {\bibinfo {author} {\bibfnamefont {J.}~\bibnamefont
  {Kurzidim}}, \bibinfo {author} {\bibfnamefont {D.}~\bibnamefont {Coslovich}},
  \ and\ \bibinfo {author} {\bibfnamefont {G.}~\bibnamefont {Kahl}},\ }\href
  {\doibase 10.1103/PhysRevLett.103.138303} {\bibfield  {journal} {\bibinfo
  {journal} {Phys. Rev. Lett.}\ }\textbf {\bibinfo {volume} {103}},\ \bibinfo
  {pages} {138303} (\bibinfo {year} {2009})}\BibitemShut {NoStop}%
\bibitem [{\citenamefont {Szamel}\ and\ \citenamefont
  {Flenner}(2013)}]{szamel2013glassy}%
  \BibitemOpen
  \bibfield  {author} {\bibinfo {author} {\bibfnamefont {G.}~\bibnamefont
  {Szamel}}\ and\ \bibinfo {author} {\bibfnamefont {E.}~\bibnamefont
  {Flenner}},\ }\href {\doibase 10.1209/0295-5075/101/66005} {\bibfield
  {journal} {\bibinfo  {journal} {Europhys. Lett.}\ }\textbf {\bibinfo {volume}
  {101}},\ \bibinfo {pages} {66005} (\bibinfo {year} {2013})}\BibitemShut
  {NoStop}%
\bibitem [{\citenamefont {Biroli}\ \emph
  {et~al.}(2021{\natexlab{b}})\citenamefont {Biroli}, \citenamefont
  {Charbonneau}, \citenamefont {Hu}, \citenamefont {Ikeda}, \citenamefont
  {Szamel},\ and\ \citenamefont {Zamponi}}]{biroli2020mean}%
  \BibitemOpen
  \bibfield  {author} {\bibinfo {author} {\bibfnamefont {G.}~\bibnamefont
  {Biroli}}, \bibinfo {author} {\bibfnamefont {P.}~\bibnamefont {Charbonneau}},
  \bibinfo {author} {\bibfnamefont {Y.}~\bibnamefont {Hu}}, \bibinfo {author}
  {\bibfnamefont {H.}~\bibnamefont {Ikeda}}, \bibinfo {author} {\bibfnamefont
  {G.}~\bibnamefont {Szamel}}, \ and\ \bibinfo {author} {\bibfnamefont
  {F.}~\bibnamefont {Zamponi}},\ }\href@noop {} {\bibfield  {journal} {\bibinfo
   {journal} {arXiv preprint}\ } (\bibinfo {year} {2021}{\natexlab{b}})},\
  \Eprint {http://arxiv.org/abs/arXiv:2102.12019} {arXiv:2102.12019}
  \BibitemShut {NoStop}%
\bibitem [{\citenamefont {Morin}\ \emph {et~al.}(2017)\citenamefont {Morin},
  \citenamefont {Cardozo}, \citenamefont {Chikkadi},\ and\ \citenamefont
  {Bartolo}}]{morin2017diffusion}%
  \BibitemOpen
  \bibfield  {author} {\bibinfo {author} {\bibfnamefont {A.}~\bibnamefont
  {Morin}}, \bibinfo {author} {\bibfnamefont {D.~L.}\ \bibnamefont {Cardozo}},
  \bibinfo {author} {\bibfnamefont {V.}~\bibnamefont {Chikkadi}}, \ and\
  \bibinfo {author} {\bibfnamefont {D.}~\bibnamefont {Bartolo}},\ }\href
  {\doibase 10.1103/PhysRevE.96.042611} {\bibfield  {journal} {\bibinfo
  {journal} {Phys. Rev. E}\ }\textbf {\bibinfo {volume} {96}},\ \bibinfo
  {pages} {042611} (\bibinfo {year} {2017})}\BibitemShut {NoStop}%
\bibitem [{\citenamefont {Petersen}\ and\ \citenamefont
  {Franosch}(2019)}]{petersen2019anomalous}%
  \BibitemOpen
  \bibfield  {author} {\bibinfo {author} {\bibfnamefont {C.~F.}\ \bibnamefont
  {Petersen}}\ and\ \bibinfo {author} {\bibfnamefont {T.}~\bibnamefont
  {Franosch}},\ }\href {\doibase 10.1039/c9sm00442d} {\bibfield  {journal}
  {\bibinfo  {journal} {Soft Matter}\ }\textbf {\bibinfo {volume} {15}},\
  \bibinfo {pages} {3906} (\bibinfo {year} {2019})}\BibitemShut {NoStop}%
\bibitem [{\citenamefont {Bauer}\ \emph {et~al.}(2010)\citenamefont {Bauer},
  \citenamefont {H{\"o}fling}, \citenamefont {Munk}, \citenamefont {Frey},\
  and\ \citenamefont {Franosch}}]{bauer2010localization}%
  \BibitemOpen
  \bibfield  {author} {\bibinfo {author} {\bibfnamefont {T.}~\bibnamefont
  {Bauer}}, \bibinfo {author} {\bibfnamefont {F.}~\bibnamefont {H{\"o}fling}},
  \bibinfo {author} {\bibfnamefont {T.}~\bibnamefont {Munk}}, \bibinfo {author}
  {\bibfnamefont {E.}~\bibnamefont {Frey}}, \ and\ \bibinfo {author}
  {\bibfnamefont {T.}~\bibnamefont {Franosch}},\ }\href {\doibase
  10.1140/epjst/e2010-01313-1} {\bibfield  {journal} {\bibinfo  {journal} {Eur.
  Phys. J Spec. Top.}\ }\textbf {\bibinfo {volume} {189}},\ \bibinfo {pages}
  {103} (\bibinfo {year} {2010})}\BibitemShut {NoStop}%
\bibitem [{\citenamefont {Spanner}\ \emph {et~al.}(2016)\citenamefont
  {Spanner}, \citenamefont {H{\"o}fling}, \citenamefont {Kapfer}, \citenamefont
  {Mecke}, \citenamefont {Schr{\"o}der-Turk},\ and\ \citenamefont
  {Franosch}}]{spanner2016splitting}%
  \BibitemOpen
  \bibfield  {author} {\bibinfo {author} {\bibfnamefont {M.}~\bibnamefont
  {Spanner}}, \bibinfo {author} {\bibfnamefont {F.}~\bibnamefont
  {H{\"o}fling}}, \bibinfo {author} {\bibfnamefont {S.~C.}\ \bibnamefont
  {Kapfer}}, \bibinfo {author} {\bibfnamefont {K.~R.}\ \bibnamefont {Mecke}},
  \bibinfo {author} {\bibfnamefont {G.~E.}\ \bibnamefont {Schr{\"o}der-Turk}},
  \ and\ \bibinfo {author} {\bibfnamefont {T.}~\bibnamefont {Franosch}},\
  }\href {\doibase 10.1103/PhysRevLett.116.060601} {\bibfield  {journal}
  {\bibinfo  {journal} {Phys. Rev. Lett.}\ }\textbf {\bibinfo {volume} {116}},\
  \bibinfo {pages} {060601} (\bibinfo {year} {2016})}\BibitemShut {NoStop}%
\bibitem [{\citenamefont {Lubensky}\ and\ \citenamefont
  {Tremblay}(1986)}]{lubensky1986varepsilon}%
  \BibitemOpen
  \bibfield  {author} {\bibinfo {author} {\bibfnamefont {T.~C.}\ \bibnamefont
  {Lubensky}}\ and\ \bibinfo {author} {\bibfnamefont {A.-M.~S.}\ \bibnamefont
  {Tremblay}},\ }\href {\doibase 10.1103/PhysRevB.34.3408} {\bibfield
  {journal} {\bibinfo  {journal} {Phys. Rev. B}\ }\textbf {\bibinfo {volume}
  {34}},\ \bibinfo {pages} {3408} (\bibinfo {year} {1986})}\BibitemShut
  {NoStop}%
\bibitem [{\citenamefont {Stenull}\ and\ \citenamefont
  {Janssen}(2001)}]{stenull2001conductivity}%
  \BibitemOpen
  \bibfield  {author} {\bibinfo {author} {\bibfnamefont {O.}~\bibnamefont
  {Stenull}}\ and\ \bibinfo {author} {\bibfnamefont {H.-K.}\ \bibnamefont
  {Janssen}},\ }\href {\doibase 10.1103/PhysRevE.64.056105} {\bibfield
  {journal} {\bibinfo  {journal} {Phys. Rev. E}\ }\textbf {\bibinfo {volume}
  {64}},\ \bibinfo {pages} {056105} (\bibinfo {year} {2001})}\BibitemShut
  {NoStop}%
\bibitem [{\citenamefont {Biroli}\ \emph {et~al.}(2019)\citenamefont {Biroli},
  \citenamefont {Charbonneau},\ and\ \citenamefont {Hu}}]{biroli2019dynamics}%
  \BibitemOpen
  \bibfield  {author} {\bibinfo {author} {\bibfnamefont {G.}~\bibnamefont
  {Biroli}}, \bibinfo {author} {\bibfnamefont {P.}~\bibnamefont {Charbonneau}},
  \ and\ \bibinfo {author} {\bibfnamefont {Y.}~\bibnamefont {Hu}},\ }\href
  {\doibase 10.1103/PhysRevE.99.022118} {\bibfield  {journal} {\bibinfo
  {journal} {Phys. Rev. E}\ }\textbf {\bibinfo {volume} {99}},\ \bibinfo
  {pages} {022118} (\bibinfo {year} {2019})}\BibitemShut {NoStop}%
\bibitem [{\citenamefont {Penrose}\ \emph {et~al.}(1996)\citenamefont {Penrose}
  \emph {et~al.}}]{penrose1996continuum}%
  \BibitemOpen
  \bibfield  {author} {\bibinfo {author} {\bibfnamefont {M.~D.}\ \bibnamefont
  {Penrose}} \emph {et~al.},\ }\href {\doibase 10.1214/aoap/1034968142}
  {\bibfield  {journal} {\bibinfo  {journal} {Ann. Appl. Probab}\ }\textbf
  {\bibinfo {volume} {6}},\ \bibinfo {pages} {528} (\bibinfo {year}
  {1996})}\BibitemShut {NoStop}%
\bibitem [{\citenamefont {Kerstein}(1983)}]{kerstein1983equivalence}%
  \BibitemOpen
  \bibfield  {author} {\bibinfo {author} {\bibfnamefont {A.~R.}\ \bibnamefont
  {Kerstein}},\ }\href {\doibase 10.1088/0305-4470/16/13/031} {\bibfield
  {journal} {\bibinfo  {journal} {J. Phys. A}\ }\textbf {\bibinfo {volume}
  {16}},\ \bibinfo {pages} {3071} (\bibinfo {year} {1983})}\BibitemShut
  {NoStop}%
\bibitem [{\citenamefont {Caroli}\ \emph {et~al.}(2019)\citenamefont {Caroli},
  \citenamefont {Pell{\'e}}, \citenamefont {Rouxel-Labb{\'e}},\ and\
  \citenamefont {Teillaud}}]{cgal:ct-pt3-19a}%
  \BibitemOpen
  \bibfield  {author} {\bibinfo {author} {\bibfnamefont {M.}~\bibnamefont
  {Caroli}}, \bibinfo {author} {\bibfnamefont {A.}~\bibnamefont {Pell{\'e}}},
  \bibinfo {author} {\bibfnamefont {M.}~\bibnamefont {Rouxel-Labb{\'e}}}, \
  and\ \bibinfo {author} {\bibfnamefont {M.}~\bibnamefont {Teillaud}},\ }in\
  \href
  {https://doc.cgal.org/4.14/Manual/packages.html#PkgPeriodic3Triangulation3}
  {\emph {\bibinfo {booktitle} {{CGAL} User and Reference Manual}}}\ (\bibinfo
  {publisher} {{CGAL Editorial Board}},\ \bibinfo {year} {2019})\ \bibinfo
  {edition} {{4.14}}\ ed.\BibitemShut {Stop}%
\bibitem [{\citenamefont {Caroli}\ and\ \citenamefont
  {Teillaud}(2011)}]{caroli2011delaunay}%
  \BibitemOpen
  \bibfield  {author} {\bibinfo {author} {\bibfnamefont {M.}~\bibnamefont
  {Caroli}}\ and\ \bibinfo {author} {\bibfnamefont {M.}~\bibnamefont
  {Teillaud}},\ }in\ \href {\doibase 10.1145/1998196.1998236} {\emph {\bibinfo
  {booktitle} {Proceedings of the twenty-seventh annual symposium on
  Computational geometry}}}\ (\bibinfo {year} {2011})\ pp.\ \bibinfo {pages}
  {274--282}\BibitemShut {NoStop}%
\bibitem [{\citenamefont {Caroli}\ and\ \citenamefont
  {Teillaud}(2016)}]{caroli2016delaunay}%
  \BibitemOpen
  \bibfield  {author} {\bibinfo {author} {\bibfnamefont {M.}~\bibnamefont
  {Caroli}}\ and\ \bibinfo {author} {\bibfnamefont {M.}~\bibnamefont
  {Teillaud}},\ }\href {\doibase 10.1007/s00454-016-9782-6} {\bibfield
  {journal} {\bibinfo  {journal} {Discrete Comput. Geom.}\ }\textbf {\bibinfo
  {volume} {55}},\ \bibinfo {pages} {827} (\bibinfo {year} {2016})}\BibitemShut
  {NoStop}%
\bibitem [{\citenamefont {Charbonneau}\ \emph {et~al.}(2013)\citenamefont
  {Charbonneau}, \citenamefont {Charbonneau},\ and\ \citenamefont
  {Tarjus}}]{charbonneau2013geometrical}%
  \BibitemOpen
  \bibfield  {author} {\bibinfo {author} {\bibfnamefont {B.}~\bibnamefont
  {Charbonneau}}, \bibinfo {author} {\bibfnamefont {P.}~\bibnamefont
  {Charbonneau}}, \ and\ \bibinfo {author} {\bibfnamefont {G.}~\bibnamefont
  {Tarjus}},\ }\href {\doibase 10.1063/1.4770498} {\bibfield  {journal}
  {\bibinfo  {journal} {J. Chem. Phys.}\ }\textbf {\bibinfo {volume} {138}},\
  \bibinfo {pages} {12A515} (\bibinfo {year} {2013})}\BibitemShut {NoStop}%
\bibitem [{\citenamefont {Morse}\ and\ \citenamefont
  {Corwin}(2014)}]{morse2014geometric}%
  \BibitemOpen
  \bibfield  {author} {\bibinfo {author} {\bibfnamefont {P.~K.}\ \bibnamefont
  {Morse}}\ and\ \bibinfo {author} {\bibfnamefont {E.~I.}\ \bibnamefont
  {Corwin}},\ }\href {\doibase 10.1103/PhysRevLett.112.115701} {\bibfield
  {journal} {\bibinfo  {journal} {Phys. Rev. Lett.}\ }\textbf {\bibinfo
  {volume} {112}},\ \bibinfo {pages} {115701} (\bibinfo {year}
  {2014})}\BibitemShut {NoStop}%
\bibitem [{\citenamefont {Tomilov}(2016)}]{Quickhull2016}%
  \BibitemOpen
  \bibfield  {author} {\bibinfo {author} {\bibfnamefont {A.~V.}\ \bibnamefont
  {Tomilov}},\ }\href {https://github.com/tomilov/quickhull} {\enquote
  {\bibinfo {title} {Header-only single-class implementation of the quickhull
  algorithm for convex hulls finding in arbitrary dimension space},}\ }
  (\bibinfo {year} {2016}),\ \bibinfo {note} {based
  on~\cite{barber1996quickhull,mehlhorn1999checking}}\BibitemShut {NoStop}%
\bibitem [{\citenamefont {Boissonnat}\ and\ \citenamefont
  {Delage}(2005)}]{boissonnat2005convex}%
  \BibitemOpen
  \bibfield  {author} {\bibinfo {author} {\bibfnamefont {J.-D.}\ \bibnamefont
  {Boissonnat}}\ and\ \bibinfo {author} {\bibfnamefont {C.}~\bibnamefont
  {Delage}},\ }in\ \href {\doibase 10.1007/11561071_34} {\emph {\bibinfo
  {booktitle} {European Symposium on Algorithms}}}\ (\bibinfo {organization}
  {Springer},\ \bibinfo {year} {2005})\ pp.\ \bibinfo {pages}
  {367--378}\BibitemShut {NoStop}%
\bibitem [{\citenamefont {Newman}\ and\ \citenamefont
  {Ziff}(2001)}]{newman2001fast}%
  \BibitemOpen
  \bibfield  {author} {\bibinfo {author} {\bibfnamefont {M.~E.~J.}\
  \bibnamefont {Newman}}\ and\ \bibinfo {author} {\bibfnamefont {R.~M.}\
  \bibnamefont {Ziff}},\ }\href {\doibase 10.1103/PhysRevE.64.016706}
  {\bibfield  {journal} {\bibinfo  {journal} {Phys. Rev. E}\ }\textbf {\bibinfo
  {volume} {64}},\ \bibinfo {pages} {016706} (\bibinfo {year}
  {2001})}\BibitemShut {NoStop}%
\bibitem [{\citenamefont {Mertens}\ and\ \citenamefont
  {Moore}(2012)}]{mertens2012continuum}%
  \BibitemOpen
  \bibfield  {author} {\bibinfo {author} {\bibfnamefont {S.}~\bibnamefont
  {Mertens}}\ and\ \bibinfo {author} {\bibfnamefont {C.}~\bibnamefont
  {Moore}},\ }\href {\doibase 10.1103/PhysRevE.86.061109} {\bibfield  {journal}
  {\bibinfo  {journal} {Phys. Rev. E}\ }\textbf {\bibinfo {volume} {86}},\
  \bibinfo {pages} {061109} (\bibinfo {year} {2012})}\BibitemShut {NoStop}%
\bibitem [{\citenamefont {Convay}\ and\ \citenamefont
  {Sloane}(1982)}]{convay1982fast}%
  \BibitemOpen
  \bibfield  {author} {\bibinfo {author} {\bibfnamefont {J.~H.}\ \bibnamefont
  {Convay}}\ and\ \bibinfo {author} {\bibfnamefont {N.~J.~A.}\ \bibnamefont
  {Sloane}},\ }\href {\doibase 10.1109/TIT.1982.1056484} {\bibfield  {journal}
  {\bibinfo  {journal} {IEEE Trans. Inf. Theory}\ }\textbf {\bibinfo {volume}
  {28}},\ \bibinfo {pages} {227} (\bibinfo {year} {1982})}\BibitemShut
  {NoStop}%
\bibitem [{\citenamefont {Charbonneau}\ \emph {et~al.}(2014)\citenamefont
  {Charbonneau}, \citenamefont {Jin}, \citenamefont {Parisi},\ and\
  \citenamefont {Zamponi}}]{charbonneau2014hopping}%
  \BibitemOpen
  \bibfield  {author} {\bibinfo {author} {\bibfnamefont {P.}~\bibnamefont
  {Charbonneau}}, \bibinfo {author} {\bibfnamefont {Y.}~\bibnamefont {Jin}},
  \bibinfo {author} {\bibfnamefont {G.}~\bibnamefont {Parisi}}, \ and\ \bibinfo
  {author} {\bibfnamefont {F.}~\bibnamefont {Zamponi}},\ }\href {\doibase
  10.1073/pnas.1417182111} {\bibfield  {journal} {\bibinfo  {journal} {Proc.
  Natl. Acad. Sci. U.S.A}\ }\textbf {\bibinfo {volume} {111}},\ \bibinfo
  {pages} {15025} (\bibinfo {year} {2014})}\BibitemShut {NoStop}%
\bibitem [{\citenamefont {Leath}(1976)}]{leath1976clustersize}%
  \BibitemOpen
  \bibfield  {author} {\bibinfo {author} {\bibfnamefont {P.~L.}\ \bibnamefont
  {Leath}},\ }\href {\doibase 10.1103/PhysRevB.14.5046} {\bibfield  {journal}
  {\bibinfo  {journal} {Phys. Rev. B}\ }\textbf {\bibinfo {volume} {14}},\
  \bibinfo {pages} {5046} (\bibinfo {year} {1976})}\BibitemShut {NoStop}%
\bibitem [{\citenamefont {Sastry}\ \emph {et~al.}(1997)\citenamefont {Sastry},
  \citenamefont {Corti}, \citenamefont {Debenedetti},\ and\ \citenamefont
  {Stillinger}}]{sastry1997statistical}%
  \BibitemOpen
  \bibfield  {author} {\bibinfo {author} {\bibfnamefont {S.}~\bibnamefont
  {Sastry}}, \bibinfo {author} {\bibfnamefont {D.~S.}\ \bibnamefont {Corti}},
  \bibinfo {author} {\bibfnamefont {P.~G.}\ \bibnamefont {Debenedetti}}, \ and\
  \bibinfo {author} {\bibfnamefont {F.~H.}\ \bibnamefont {Stillinger}},\ }\href
  {\doibase 10.1103/PhysRevE.56.5524} {\bibfield  {journal} {\bibinfo
  {journal} {Phys. Rev. E}\ }\textbf {\bibinfo {volume} {56}},\ \bibinfo
  {pages} {5524} (\bibinfo {year} {1997})}\BibitemShut {NoStop}%
\bibitem [{\citenamefont {Devillers}\ \emph {et~al.}(2019)\citenamefont
  {Devillers}, \citenamefont {Hornus},\ and\ \citenamefont
  {Jamin}}]{cgal:hdj-t-19a}%
  \BibitemOpen
  \bibfield  {author} {\bibinfo {author} {\bibfnamefont {O.}~\bibnamefont
  {Devillers}}, \bibinfo {author} {\bibfnamefont {S.}~\bibnamefont {Hornus}}, \
  and\ \bibinfo {author} {\bibfnamefont {C.}~\bibnamefont {Jamin}},\ }in\ \href
  {https://doc.cgal.org/4.14/Manual/packages.html#PkgTriangulations} {\emph
  {\bibinfo {booktitle} {{CGAL} User and Reference Manual}}}\ (\bibinfo
  {publisher} {{CGAL Editorial Board}},\ \bibinfo {year} {2019})\ \bibinfo
  {edition} {{4.14}}\ ed.\BibitemShut {Stop}%
\bibitem [{\citenamefont {Devroye}(1986)}]{devroye1986nonuniform}%
  \BibitemOpen
  \bibfield  {author} {\bibinfo {author} {\bibfnamefont {L.}~\bibnamefont
  {Devroye}},\ }\href {\doibase 10.1007/978-1-4613-8643-8} {\emph {\bibinfo
  {title} {Non-uniform random variate generation}}}\ (\bibinfo  {publisher}
  {Springer-Verlag},\ \bibinfo {address} {New York, New York 10010, USA},\
  \bibinfo {year} {1986})\BibitemShut {NoStop}%
\bibitem [{\citenamefont {Newman}\ and\ \citenamefont
  {Ziff}(2000)}]{newman2000efficient}%
  \BibitemOpen
  \bibfield  {author} {\bibinfo {author} {\bibfnamefont {M.~E.~J.}\
  \bibnamefont {Newman}}\ and\ \bibinfo {author} {\bibfnamefont {R.~M.}\
  \bibnamefont {Ziff}},\ }\href {\doibase 10.1103/PhysRevLett.85.4104}
  {\bibfield  {journal} {\bibinfo  {journal} {Phys. Rev. Lett.}\ }\textbf
  {\bibinfo {volume} {85}},\ \bibinfo {pages} {4104} (\bibinfo {year}
  {2000})}\BibitemShut {NoStop}%
\bibitem [{\citenamefont {Yi}\ and\ \citenamefont
  {Esmail}(2012)}]{yi2012computational}%
  \BibitemOpen
  \bibfield  {author} {\bibinfo {author} {\bibfnamefont {Y.}~\bibnamefont
  {Yi}}\ and\ \bibinfo {author} {\bibfnamefont {K.}~\bibnamefont {Esmail}},\
  }\href {\doibase 10.1063/1.4730333} {\bibfield  {journal} {\bibinfo
  {journal} {J. Appl. Phys.}\ }\textbf {\bibinfo {volume} {111}},\ \bibinfo
  {pages} {124903} (\bibinfo {year} {2012})}\BibitemShut {NoStop}%
\bibitem [{\citenamefont {Koza}\ and\ \citenamefont
  {Po{\l}a}(2016)}]{koza2016discrete}%
  \BibitemOpen
  \bibfield  {author} {\bibinfo {author} {\bibfnamefont {Z.}~\bibnamefont
  {Koza}}\ and\ \bibinfo {author} {\bibfnamefont {J.}~\bibnamefont {Po{\l}a}},\
  }\href {\doibase 10.1088/1742-5468/2016/10/103206} {\bibfield  {journal}
  {\bibinfo  {journal} {J. Stat. Mech. Theory Exp.}\ }\textbf {\bibinfo
  {volume} {2016}},\ \bibinfo {pages} {103206} (\bibinfo {year}
  {2016})}\BibitemShut {NoStop}%
\bibitem [{\citenamefont {Grassberger}(1999)}]{grassberger1999conductivity}%
  \BibitemOpen
  \bibfield  {author} {\bibinfo {author} {\bibfnamefont {P.}~\bibnamefont
  {Grassberger}},\ }\href {\doibase 10.1016/S0378-4371(98)00435-X} {\bibfield
  {journal} {\bibinfo  {journal} {Physica A}\ }\textbf {\bibinfo {volume}
  {262}},\ \bibinfo {pages} {251} (\bibinfo {year} {1999})}\BibitemShut
  {NoStop}%
\bibitem [{\citenamefont {Berthier}\ and\ \citenamefont
  {Biroli}(2011)}]{berthier2011theoretical}%
  \BibitemOpen
  \bibfield  {author} {\bibinfo {author} {\bibfnamefont {L.}~\bibnamefont
  {Berthier}}\ and\ \bibinfo {author} {\bibfnamefont {G.}~\bibnamefont
  {Biroli}},\ }\href {\doibase 10.1103/RevModPhys.83.587} {\bibfield  {journal}
  {\bibinfo  {journal} {Rev. Mod. Phys.}\ }\textbf {\bibinfo {volume} {83}},\
  \bibinfo {pages} {587} (\bibinfo {year} {2011})}\BibitemShut {NoStop}%
\bibitem [{\citenamefont {Pordes}\ \emph {et~al.}(2007)\citenamefont {Pordes},
  \citenamefont {Petravick}, \citenamefont {Kramer}, \citenamefont {Olson},
  \citenamefont {Livny}, \citenamefont {Roy}, \citenamefont {Avery},
  \citenamefont {Blackburn}, \citenamefont {Wenaus}, \citenamefont
  {W{\"u}rthwein}, \citenamefont {Foster}, \citenamefont {Gardner},
  \citenamefont {Wilde}, \citenamefont {Blatecky}, \citenamefont {McGee},\ and\
  \citenamefont {Quick}}]{osg07}%
  \BibitemOpen
  \bibfield  {author} {\bibinfo {author} {\bibfnamefont {R.}~\bibnamefont
  {Pordes}}, \bibinfo {author} {\bibfnamefont {D.}~\bibnamefont {Petravick}},
  \bibinfo {author} {\bibfnamefont {B.}~\bibnamefont {Kramer}}, \bibinfo
  {author} {\bibfnamefont {D.}~\bibnamefont {Olson}}, \bibinfo {author}
  {\bibfnamefont {M.}~\bibnamefont {Livny}}, \bibinfo {author} {\bibfnamefont
  {A.}~\bibnamefont {Roy}}, \bibinfo {author} {\bibfnamefont {P.}~\bibnamefont
  {Avery}}, \bibinfo {author} {\bibfnamefont {K.}~\bibnamefont {Blackburn}},
  \bibinfo {author} {\bibfnamefont {T.}~\bibnamefont {Wenaus}}, \bibinfo
  {author} {\bibfnamefont {F.}~\bibnamefont {W{\"u}rthwein}}, \bibinfo {author}
  {\bibfnamefont {I.}~\bibnamefont {Foster}}, \bibinfo {author} {\bibfnamefont
  {R.}~\bibnamefont {Gardner}}, \bibinfo {author} {\bibfnamefont
  {M.}~\bibnamefont {Wilde}}, \bibinfo {author} {\bibfnamefont
  {A.}~\bibnamefont {Blatecky}}, \bibinfo {author} {\bibfnamefont
  {J.}~\bibnamefont {McGee}}, \ and\ \bibinfo {author} {\bibfnamefont
  {R.}~\bibnamefont {Quick}},\ }in\ \href {\doibase
  10.1088/1742-6596/78/1/012057} {\emph {\bibinfo {booktitle} {J. Phys. Conf.
  Ser.}}},\ Vol.~\bibinfo {volume} {78}\ (\bibinfo {year} {2007})\ p.\ \bibinfo
  {pages} {012057}\BibitemShut {NoStop}%
\bibitem [{\citenamefont {Sfiligoi}\ \emph {et~al.}(2009)\citenamefont
  {Sfiligoi}, \citenamefont {Bradley}, \citenamefont {Holzman}, \citenamefont
  {Mhashilkar}, \citenamefont {Padhi},\ and\ \citenamefont
  {Wurthwein}}]{osg09}%
  \BibitemOpen
  \bibfield  {author} {\bibinfo {author} {\bibfnamefont {I.}~\bibnamefont
  {Sfiligoi}}, \bibinfo {author} {\bibfnamefont {D.~C.}\ \bibnamefont
  {Bradley}}, \bibinfo {author} {\bibfnamefont {B.}~\bibnamefont {Holzman}},
  \bibinfo {author} {\bibfnamefont {P.}~\bibnamefont {Mhashilkar}}, \bibinfo
  {author} {\bibfnamefont {S.}~\bibnamefont {Padhi}}, \ and\ \bibinfo {author}
  {\bibfnamefont {F.}~\bibnamefont {Wurthwein}},\ }in\ \href {\doibase
  10.1109/CSIE.2009.950} {\emph {\bibinfo {booktitle} {2009 WRI World Congress
  on Computer Science and Information Engineering}}},\ \bibinfo {series} {2},
  Vol.~\bibinfo {volume} {2}\ (\bibinfo {year} {2009})\ pp.\ \bibinfo {pages}
  {428--432}\BibitemShut {NoStop}%
\bibitem [{lpd()}]{lpdata}%
  \BibitemOpen
  \href@noop {} {\enquote {\bibinfo {title} {Duke digital repository},}\
  }\bibinfo {howpublished} {https://doi.org/10.7924/xxxxxxxxx}\BibitemShut
  {NoStop}%
\bibitem [{\citenamefont {Li}(2011)}]{li2011concise}%
  \BibitemOpen
  \bibfield  {author} {\bibinfo {author} {\bibfnamefont {S.}~\bibnamefont
  {Li}},\ }\href {\doibase 10.3923/ajms.2011.66.70} {\bibfield  {journal}
  {\bibinfo  {journal} {Asian J. Math. Stat. Phys.}\ }\textbf {\bibinfo
  {volume} {4}},\ \bibinfo {pages} {66} (\bibinfo {year} {2011})}\BibitemShut
  {NoStop}%
\bibitem [{\citenamefont {Nemes}\ \emph {et~al.}(2016)\citenamefont {Nemes},
  \citenamefont {Olde~Daalhuis} \emph {et~al.}}]{nemes2016uniform}%
  \BibitemOpen
  \bibfield  {author} {\bibinfo {author} {\bibfnamefont {G.}~\bibnamefont
  {Nemes}}, \bibinfo {author} {\bibfnamefont {A.~B.}\ \bibnamefont
  {Olde~Daalhuis}},  \emph {et~al.},\ }\href {\doibase 10.3842/SIGMA.2016.101}
  {\bibfield  {journal} {\bibinfo  {journal} {SIGMA}\ }\textbf {\bibinfo
  {volume} {12}},\ \bibinfo {pages} {101} (\bibinfo {year} {2016})}\BibitemShut
  {NoStop}%
\bibitem [{\citenamefont {Barber}\ \emph {et~al.}(1996)\citenamefont {Barber},
  \citenamefont {Dobkin}, \citenamefont {Dobkin},\ and\ \citenamefont
  {Huhdanpaa}}]{barber1996quickhull}%
  \BibitemOpen
  \bibfield  {author} {\bibinfo {author} {\bibfnamefont {C.~B.}\ \bibnamefont
  {Barber}}, \bibinfo {author} {\bibfnamefont {D.~P.}\ \bibnamefont {Dobkin}},
  \bibinfo {author} {\bibfnamefont {D.~P.}\ \bibnamefont {Dobkin}}, \ and\
  \bibinfo {author} {\bibfnamefont {H.}~\bibnamefont {Huhdanpaa}},\ }\href
  {\doibase 10.1145/235815.235821} {\bibfield  {journal} {\bibinfo  {journal}
  {ACM Trans. Math. Softw.}\ }\textbf {\bibinfo {volume} {22}},\ \bibinfo
  {pages} {469} (\bibinfo {year} {1996})}\BibitemShut {NoStop}%
\bibitem [{\citenamefont {Mehlhorn}\ \emph {et~al.}(1999)\citenamefont
  {Mehlhorn}, \citenamefont {N{\"a}her}, \citenamefont {Seel}, \citenamefont
  {Seidel}, \citenamefont {Schilz}, \citenamefont {Schirra},\ and\
  \citenamefont {Uhrig}}]{mehlhorn1999checking}%
  \BibitemOpen
  \bibfield  {author} {\bibinfo {author} {\bibfnamefont {K.}~\bibnamefont
  {Mehlhorn}}, \bibinfo {author} {\bibfnamefont {S.}~\bibnamefont {N{\"a}her}},
  \bibinfo {author} {\bibfnamefont {M.}~\bibnamefont {Seel}}, \bibinfo {author}
  {\bibfnamefont {R.}~\bibnamefont {Seidel}}, \bibinfo {author} {\bibfnamefont
  {T.}~\bibnamefont {Schilz}}, \bibinfo {author} {\bibfnamefont
  {S.}~\bibnamefont {Schirra}}, \ and\ \bibinfo {author} {\bibfnamefont
  {C.}~\bibnamefont {Uhrig}},\ }\href {\doibase 10.1016/S0925-7721(98)00036-4}
  {\bibfield  {journal} {\bibinfo  {journal} {Comput. Geom}\ }\textbf {\bibinfo
  {volume} {12}},\ \bibinfo {pages} {85} (\bibinfo {year} {1999})}\BibitemShut
  {NoStop}%
\end{thebibliography}%

\end{document}